# Chalcogenide Glass-on-Graphene Photonics

**Hongtao Lin[1†,*], Yi Song[2†], Yizhong Huang[1,3†], Derek Kita[1†], Kaiqi Wang[1], Lan Li[1], Junying Li[1,4], Hanyu Zheng[1], Skylar Deckoff-Jones[1], Zhengqian Luo[1,3], Haozhe Wang[2], Spencer Novak[5], Anupama Yadav[5], Chung-Che Huang[6], Tian Gu[1], Daniel Hewak[6], Kathleen Richardson[5], Jing Kong[2], Juejun Hu[1,*]**

[1]*Department of Materials Science & Engineering, Massachusetts Institute of Technology, Cambridge, USA*
[2]*Department of Electrical Engineering & Computer Science, Massachusetts Institute of Technology, Cambridge, USA*
[3]*Department of Electronic Engineering, Xiamen University, Xiamen, China*
[4]*Key Laboratory of Optoelectronic Technology & System, Education Ministry of China, Chongqing University, Chongqing, China*
[5]*The College of Optics & Photonics, University of Central Florida, Orlando, USA*
[6]*Optoelectronics Research Centre, University of Southampton, Southampton, UK*

† These authors contributed equally to this work.
*hometown@mit.edu, hujuejun@mit.edu

**Abstract**

Two-dimensional (2-D) materials are of tremendous interest to integrated photonics given their singular optical characteristics spanning light emission, modulation, saturable absorption, and nonlinear optics. To harness their optical properties, these atomically thin materials are usually attached onto prefabricated devices via a transfer process. In this paper, we present a new route for 2-D material integration with planar photonics. Central to this approach is the use of chalcogenide glass, a multifunctional material which can be directly deposited and patterned on a wide variety of 2-D materials and can simultaneously function as the light guiding medium, a gate dielectric, and a passivation layer for 2-D materials. Besides claiming improved fabrication yield and throughput compared to the traditional transfer process, our technique also enables unconventional multilayer device geometries optimally designed for enhancing light-matter interactions in the 2-D layers. Capitalizing on this facile integration method, we demonstrate a series of high-performance glass-on-graphene devices including ultra-broadband on-chip polarizers, energy-efficient thermo-optic switches, as well as graphene-based mid-infrared (mid-IR) waveguide-integrated photodetectors and modulators.

The isolation of single-layer graphene in 2004 has triggered intensive investigations into 2-D crystals consisting of one or a few monolayers of atoms. With their remarkable optical properties, these materials have garnered enormous interest for their photonic applications as light emitters[1], modulators[2,3], photodetectors[4,5], saturable absorbers[6], and plasmonic sensors[7]. On-chip integration of 2-D materials with photonic devices generally relies on layer transfer, where exfoliated or delaminated 2-D membranes are attached onto prefabricated devices[8]. Despite its widespread implementation, the transfer approach has its limitations. When transferring these atomically thin crystals onto a substrate with uneven topology, the 2-D materials tend to rupture at the pattern step edges. To circumvent such damage, an additional planarization step is often mandated prior to 2-D material transfer, which complicates the process[9-12]. Further, the transferred 2-D layer resides on top of the pre-patterned devices and thus only interacts with the optical mode through the relatively weak evanescent waves.

To resolve these issues, an alternative 2-D material integration route entails growing an optically thick (comparable to optical wavelength in the medium) film directly on 2-D materials and lithographically patterning it into functional photonic devices. Besides improved processing yield and throughput compared to the traditional transfer process, this “monolithic” approach also offers several critical advantages: it enables accurate alignment of photonic components with 2-D material structures (e.g., in-plane heterojunctions) with lithographic precision, which is difficult to attain using transfer; it allows flexible placement of 2-D material layers inside a photonic structure to maximally enhance light-matter interactions; and last but not least, it heralds a truly monolithic, wafer-scale integration process with 2-D material systems where catalyst-free, large-area continuous growth on semiconductor or dielectric substrates has been realized (e.g., graphene on SiC[13], $MoS_2$ and $MoTe_2$ on $SiO_2$/Si[14,15]).

Growth of optically thick dielectric films on 2-D materials, however, is not a trivial task. Integration on graphene, the archetypal 2-D material, epitomizes the challenge. Graphene has a chemically inert surface which makes nucleation and growth of a uniform dielectric film on its surface difficult[16]. Surface modification using ozone[17], $NO_2$[18], or perylene tetracarboxylic acid[19] catalyzes nucleation, albeit at the expense of carrier mobility in graphene. Atomic Layer Deposition (ALD) has been widely adopted for gate dielectric deposition on graphene[20]; however, growing an optically thick layer using ALD is impractical. Alternatively, plasma-enhanced chemical vapor deposition (PECVD) has been attempted for silicon nitride coating on graphene, although the process requires low-density, low-power plasma with reduced deposition rate to mitigate plasma damage to graphene surface[21]. Recently, a simple spin-coating process was devised for direct polymer waveguide modulator fabrication on graphene[22]. Nevertheless, the large modal area in low-index-contrast polymer waveguides limits the resulting device footprint and performance. For other 2-D materials, especially the less stable ones such as black phosphorous[23], protection of the material’s structural integrity from high temperatures, plasma, and reactive chemicals imposes additional constraints on the integration process.

In this paper, we present a generic route for photonic integration of 2-D materials using chalcogenide glass (ChG) as the backbone optical material. Chalcogenide glasses, namely the amorphous compounds containing S, Se, and/or Te, are emerging photonic materials known for their broadband transparency, high and continuously tunable refractive indices ($n$ ~ 2 to 3.5), and large Kerr nonlinearity[24,25]. In addition to their exceptional optical properties, ChG’s are also uniquely poised for 2-D material integration. These glasses can be deposited at high rates exceeding 100 nm/min via simple single-source thermal evaporation with the substrate held near room temperature[26]. Combined with their amorphous nature and good van der Waals adhesion to different substrates without surface modification, the extremely low thermal budget allows epitaxy-free ChG coating with minimal thermal and structural damage to the substrate. Capitalizing on this "substrate-agnostic" integration capacity, prior work from our groups as well as others have demonstrated ChG integration with polymers to enable mechanically flexible photonic circuits and fibers[27,28], with infrared crystals for on-chip mid-infrared sensing[29,30], and with stacked solar cells as a high-index adhesive to minimize Fresnel reflection and achieve a then-record cell efficiency of 43.9%[31]. Here we show that ChG's can be deposited on a wide variety of 2-D materials without disrupting their structure and optoelectronic properties. Figure 1a displays the Raman spectra of monolayer graphene synthesized using chemical vapor deposition (CVD) before and after coating with a 450 nm thick thermally evaporated $Ge_{23}Sb_7S_{70}$ ChG film. No defect-related peaks (D, D' or D+G) were observed after ChG deposition, indicating that the low-temperature glass deposition does not introduce structural defects into graphene[32]. We further confirm that the structures of other 2-D materials ($MoS_2$, black phosphorus, InSe, and hexagonal

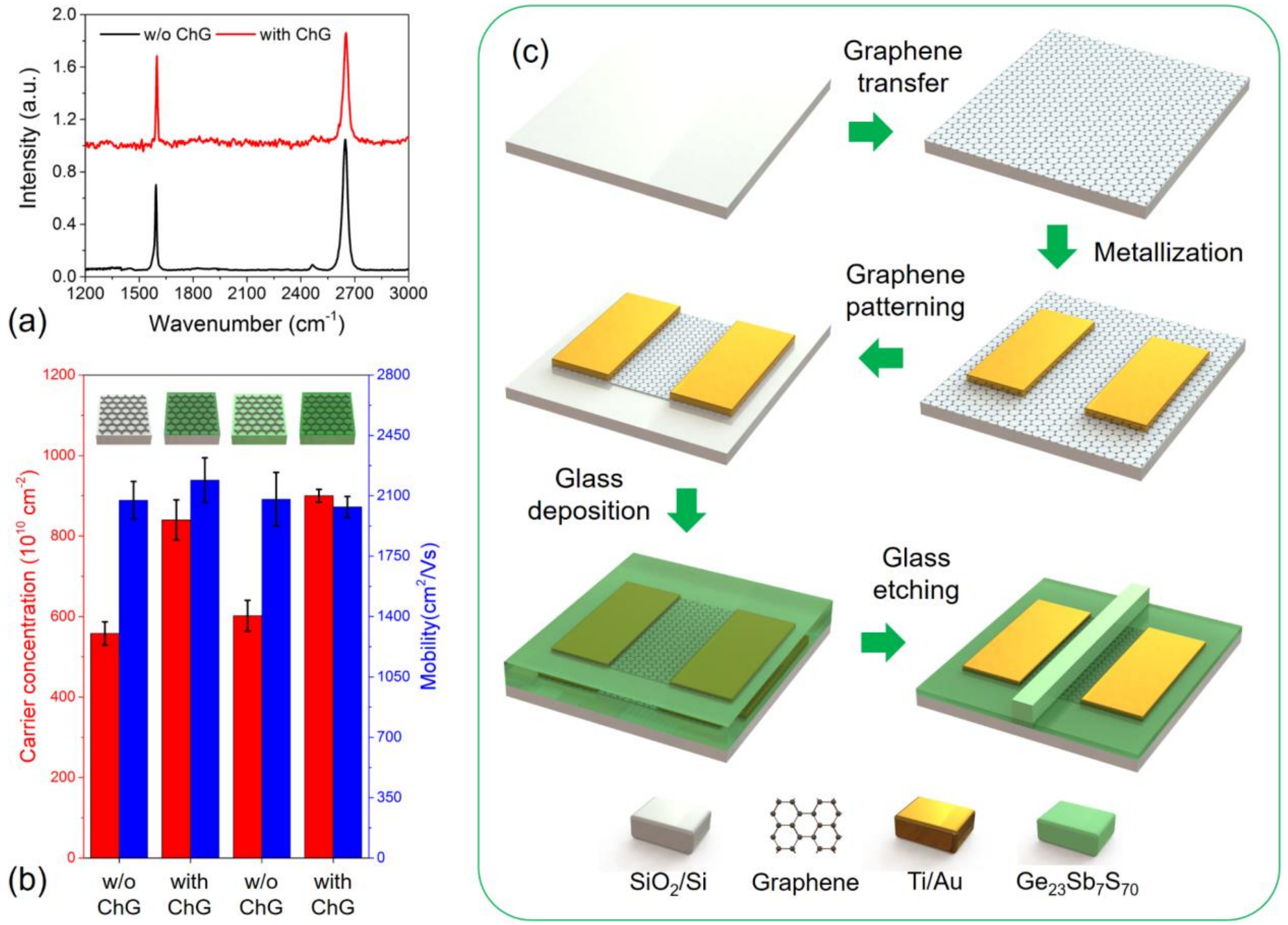

**Fig. 1. (a) Raman spectra of as-transferred monolayer CVD graphene (black) and graphene covered with a $Ge_{23}Sb_7S_{70}$ glass layer (red). Background Raman signal from the $Ge_{23}Sb_7S_{70}$ glass film has been subtracted. (b) Hall carrier concentration and mobility measured in graphene (results averaged over five samples of each type). From left to right: graphene transferred onto an oxidized silicon wafer; graphene transferred onto an oxidized silicon wafer and then covered with a $Ge_{23}Sb_7S_{70}$ glass film; graphene transferred onto a $Ge_{23}Sb_7S_{70}$ glass film; graphene sandwiched between two $Ge_{23}Sb_7S_{70}$ glass layers. (c) Schematic fabrication process flow to integrate chalcogenide glass photonic devices with graphene.**

BN) likewise remain intact after ChG deposition (Supplementary Section I). Such integration compatibility facilitates the fabrication of unconventional multi-layer structures incorporating 2-D materials to optimally engineer their interactions with the optical mode. As an example, we exploit the giant optical anisotropy of graphene and modal symmetry in graphene-sandwiched waveguides to demonstrate an ultra-broadband polarizer and a thermo-optic switch with energy efficiency an order of magnitude higher compared to previous reports.

In addition to being an optical guiding medium, the insulating $Ge_{23}Sb_7S_{70}$ glass can function as a gate dielectric and as an effective passivation barrier to prevent 2-D materials from degradation inflicted by ambient air, moisture, or corrosive chemicals (Supplementary Section II). Figure 1b evaluates the impact of $Ge_{23}Sb_7S_{70}$ glass deposition on transport properties of monolayer CVD graphene transferred onto an oxidized silicon wafer or a $Ge_{23}Sb_7S_{70}$ film on silicon. Notably, despite the increased p-doping (which normally reduces mobility), carrier mobility in graphene remains unchanged after ChG encapsulation, in contrast to most other deposited dielectrics which tend to degrade carrier mobility due to surface damage during deposition and hence increased defect density[33]. In this paper, we harness this feature to demonstrate the first mid-IR graphene waveguide modulator, where the multifunctional ChG material serves simultaneously as the waveguide and as a gate dielectric to electrostatically modulate the Fermi level in graphene.

Figure 1c illustrates the baseline fabrication protocols for the ChG-on-graphene photonic devices. Details of the fabrication process are furnished in Methods. The following sections present four classes of novel devices leveraging the new integration strategy to reap unique performance benefits. We note that while the devices described herein were fabricated using the specific combination of thermally evaporated $Ge_{23}Sb_7S_{70}$ glass and graphene, we have validated the integration process based on other 2-D materials and ChG compositions formed using alternative methods including solution processing and nanoimprint[34] (Supplementary Section III). The ChG/2D material integration process is therefore generic and can be adapted to meet diverse device design and application needs.

**Ultra-broadband on-chip waveguide polarizer**

Unlike traditional graphene-integrated devices where the transferred graphene layer is located outside the waveguide core, here we introduce a new multilayer waveguide platform comprising a graphene monolayer situated at the center of a symmetrically cladded strip waveguide (Fig. 2a). Figure 2c shows a scanning electron microscopy (SEM) image of a fabricated waveguide where a graphene film is sandwiched between two $Ge_{23}Sb_7S_{70}$ layers of equal thickness. The waveguide behaves as a polarizer as a result of the large optical anisotropy of graphene and the polarization-dependent symmetric properties of waveguide modes. To illustrate its working principle, Fig. 2b depicts the electric field components of the fundamental TM (transverse magnetic) and TE (transverse electric) modes supported in the waveguide at 1550 nm wavelength. For the TM polarization, its in-plane electric field components ($E_x$ and $E_z$) are anti-symmetric with respect to the center plane and thus vanish at the graphene layer. Since graphene acts as an optically absorbing metal in-plane and as a lossless dielectric along the out-of-plane direction[35], the waveguide becomes transparent to the TM mode. In contrast, both in-plane electric field components of the TE mode reach maximum at the waveguide center, leading to strong optical attenuation. Using experimental Fermi level data from Hall measurements, we modeled the propagation losses for the TM and TE modes as (0 – 1.5) dB/cm and (575 ± 1.5) dB/cm respectively at 1550 nm wavelength, where the error bars take into account glass thickness deviations based on realistic fabrication tolerances (Supplementary Section IV).

To precisely quantify the large polarization-dependent losses in the waveguide, we employed two device structures: ring resonators to characterize the low-loss TM mode, and unbalanced Mach Zehnder interferometers (MZI) to gauge the much higher TE-mode loss. Protocols of loss extraction are summarized in Supplementary Section V. Figures 2d and 2e plot exemplary transmission spectra of ring resonators without and with the embedded graphene layer. While TM-mode resonances are clearly visible for both types of devices, the TE-mode resonances disappear in the graphene-sandwiched waveguide, signaling significant TE polarization-selective absorption by graphene. Using the classical coupled-wave transfer matrix formalism, we calculated the excess TM-mode loss induced by graphene to be 20 dB/cm at 1550 nm, which we attribute to unevenness of graphene caused by polymer residues from the transfer process (Supplementary Section IV). The TE-mode loss was assessed based on the unbalanced MZI transmission spectra in Fig. 2f, where the extinction ratio (ER) of the transmittance undulation correlates with the differential optical attenuation induced by graphene embedded in the MZI arms. Figure 2g plots the calculated differential TE-mode absorption by graphene as a function of embedded graphene length difference in the two arms, from which we infer a TE-mode loss of 590 dB/cm near 1550 nm, which agrees well with our theoretical predictions. The results correspond to 23 dB ER and 0.8 dB insertion loss in a 400-μm-long polarizer device operating at 1550 nm wavelength, and a large figure of merit (defined as the ratio of ER to insertion loss) of 29.

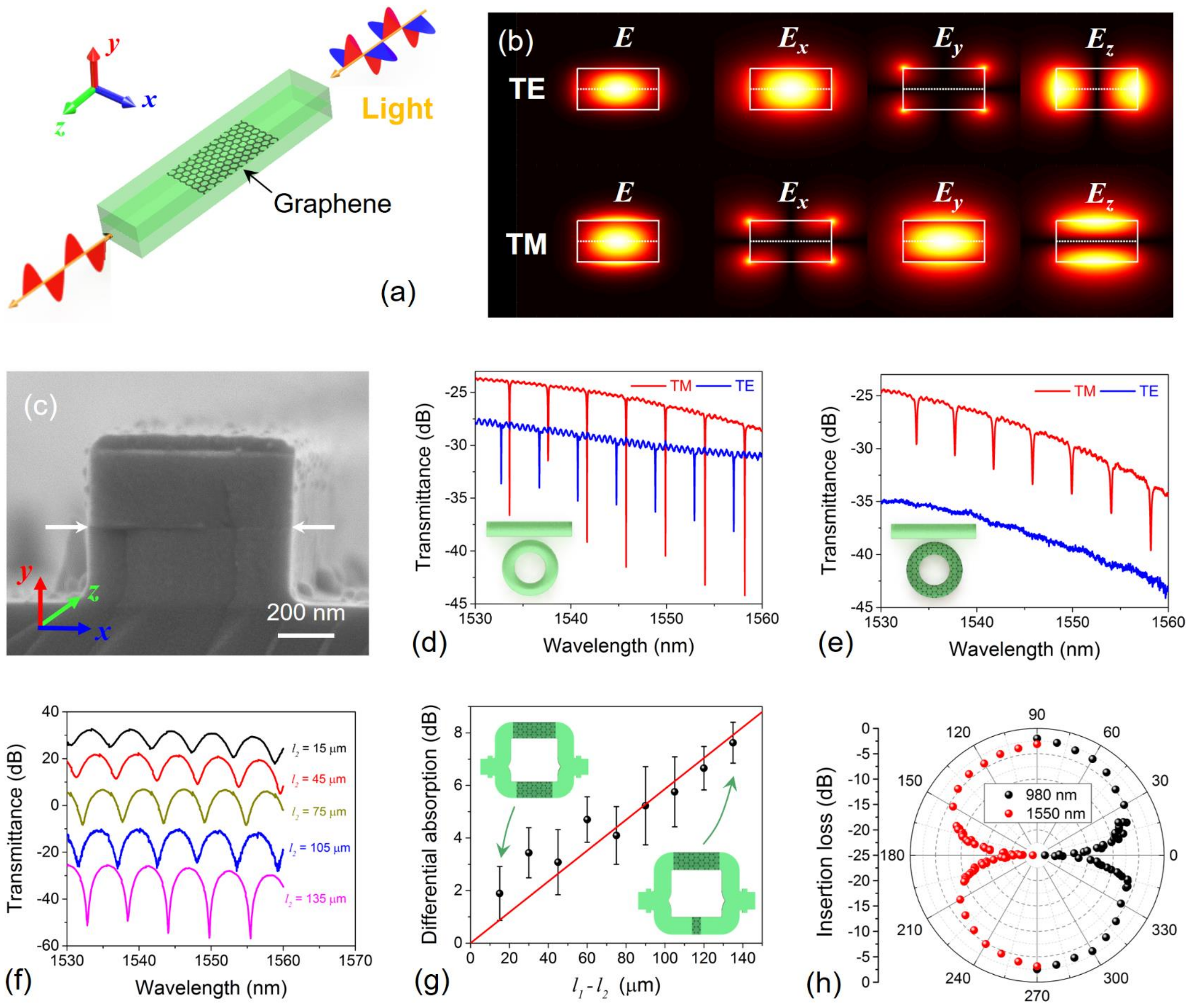

**Fig. 2. (a) Schematic diagram showing the graphene-sandwiched waveguide polarizer. (b) Simulated modal intensity and electric field component profiles for fundamental TE and TM modes in the graphene-sandwiched waveguide. (c) SEM image of the fabricated polarizer cross-section: the white arrows mark graphene layer location. (d, e) Optical transmittance through micro-ring resonators (d) without embedded graphene layer; and (e) with embedded graphene layer in the waveguide center. (f) Transmittance spectra of unbalanced MZIs where graphene strips of different lengths ($l_1$ and $l_2$) are embedded inside their two arms. For the devices displayed here, $l_1$ = 150 μm and $l_2$ is varied from 15 μm to 135 μm. (g) Differential absorption induced by graphene as a function of embedded graphene strip length difference $l_1$ - $l_2$ in two MZI arms. The error bars correspond to standard deviations of measurements performed on at least 10 devices at each $l_1$ - $l_2$ value. (h) Polar diagram showing the polarizer performance at 980 nm and 1550 nm wavelengths. The polar angle represents the angle between the polarization plane of input wave and the substrate.**

Importantly, since the polarizer design operates on material anisotropy and modal symmetry, both of which are wavelength-independent, the device is broadband in nature. To demonstrate broadband operation of the polarizer, a 400 μm-long polarizer device was characterized at 980 nm and 1550 nm wavelengths and the results are summarized in the polar diagram in Fig. 2h. The measurement procedures are elaborated in Supplementary Section VI. Consistent with the experimental results, our theoretical model confirms that the same device can operate over the broad spectral range from 940 nm to 1600 nm with a polarization extinction ratio exceeding 20 dB, which represents the largest operation bandwidth for on-chip waveguide polarizers (Supplementary Section VI).

**Energy-efficient photonic crystal thermo-optic switch**

The TM-transparent sandwich waveguide provides an example where graphene is embedded inside a waveguide without incurring excess optical loss. This counterintuitive observation opens up the application of graphene as a broadband transparent conductor. In the following we apply the embedded graphene electrode as resistive heaters to realize a thermo-optic switch with unprecedented energy efficiency. Unlike traditional metal heaters which have to be placed several microns away from the waveguide to suppress parasitic optical absorption, the waveguide-integrated graphene heater offers superior energy efficiency because of the much smaller thermal mass and large spatial overlap of the optical mode with the heating zone.

Figure 3d schematically illustrates the device structure consisting of a waveguide-coupled photonic crystal nanobeam cavity formed through depth modulation of side Bragg gratings[36]. A graphene monolayer is embedded in the center of the nanobeam cavity waveguide and connected to a pair of electrodes as described in Supplementary Section VII. Figure 3a shows a top-view SEM micrograph of the graphene-embedded nanobeam, which supports a single resonant mode near 1570 nm (Fig. 3b). When a bias voltage is applied across the electrodes, the graphene and the cavity are resistively heated, leading to a thermo-optic spectral drift of the cavity resonance. Figure 3c depicts the simulated temperature profile as a result of resistive heating in graphene. Since the graphene conductor is placed directly inside the waveguide core, this unique geometry leads to strong thermal confinement and large spatial overlap between the heating zone and the cavity mode, both of which contribute to improved energy efficiency. Figure 3e presents the transmission spectra of the cavity showing progressive resonance detuning with increasing input power. As is shown in Fig. 3f, the measured resonance shift agrees well with our finite element modeling (Supplementary Section VIII). The slope of the curve indicates a record energy efficiency of 10 nm/mW, which represents almost an order of magnitude improvement compared to the best values previously reported in on-chip thermo-optic switches and tuning devices[37].

To elucidate the device physics underlying the exceptional energy efficiency, we analyzed the switch's performance characteristics using a lumped element model (Supplementary Section IX). A figure of merit for thermo-optic switches, defined as the inverse of the product of rise time and power consumption, is often cited when drawing comparison between different technologies[38].

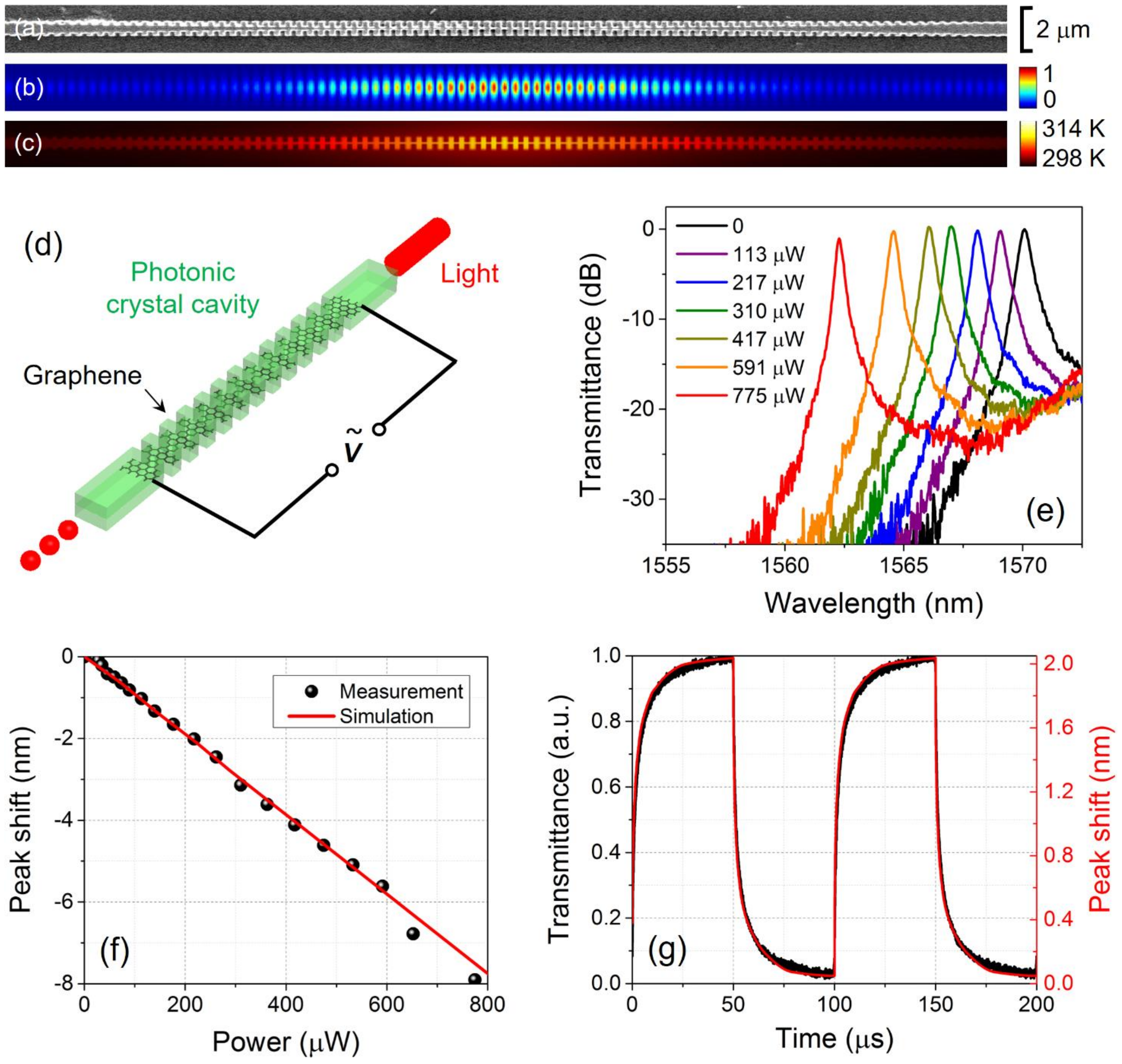

**Fig. 3. (a) Top-view SEM micrograph of the photonic crystal thermo-optic switch. (b) Simulated optical mode profile in the photonic crystal cavity. (c) Temperature distribution in the device when electric current is flowing through the embedded graphene heater. (d) Schematic illustration of the thermo-optic switch structure, which consists of a graphene layer embedded in the center of a photonic crystal nanobeam cavity. (e) Optical transmission spectra of the switch at varying input power levels into the graphene heater. (f) Thermo-optic resonant wavelength shift: the solid line represents FEM simulation results whereas the dots are experimental data. (g) Time-domain response of the switch to a square-wave driving current at 10 kHz.**

With a low switching energy of 0.11 mW and a 10%-to-90% rise time of 14 μs (Fig. 3g), our device features a FOM of 0.65 $mW^{-1}·μs^{-1}$, which is among the highest values reported in an on-chip thermo-optic switch (Supplementary Section X).

## Mid-IR waveguide-integrated photodetector

Our integration scheme equally applies to optoelectronic devices where graphene becomes the active medium. The broadband infrared transparency of ChG's makes them particularly appealing for integration with graphene, whose zero-gap nature potentially enables broadband optical detection. Our approach simplifies the graphene detector and waveguide integration process through direct deposition and patterning of ChG waveguides and metal contacts on monolayer CVD graphene (Fig. 4a inset). Figure 4a shows a tilted view of the fabricated detector. The detector operates in a photothermoelectric (PTE) mode and thus the device assumes an asymmetric configuration where the waveguide is intentionally offset from the center line between the metal electrodes. The device was characterized by launching TE-polarized light from a mid-IR laser into the waveguide. The PTE mechanism also explains the non-vanishing photoresponse at zero bias (Fig. 4b), which corroborates that bolometric effect is not a main contributor to photocurrent. As shown in Fig. 4c, responsivity of the detector increases with bias voltage due to bias-induced spatial re-distribution of carriers in graphene, consistent with the PTE mechanism[39]. The device exhibits broadband photoresponse over the entire scanning range of our tunable laser (2.0 – 2.55 μm) with a peak responsivity of 250 mA/W at 2.03 μm wavelength (Fig. 4d). The responsivity figure is on par with, or in some cases superior to, state-of-the-art waveguide-integrated graphene detectors operating in the mid-IR[40] and near-IR[9,11,39,41]. Hall measurements indicate that the Fermi level of graphene used in the device is located at 0.34 eV below the Dirac point owing to substrate doping[42]. Consequently, the reduced responsivity observed at longer wavelengths manifests the onset of Pauli blocking and decreased optical absorption in the p-type graphene. We have modeled the wavelength-dependent absorption in graphene (Supplementary Section XI) and the predicted wavelength scaling of graphene absorption is plotted in Fig. 4d. The agreement between the calculated graphene absorption spectrum and the measured responsivity trend validates the hypothesis.

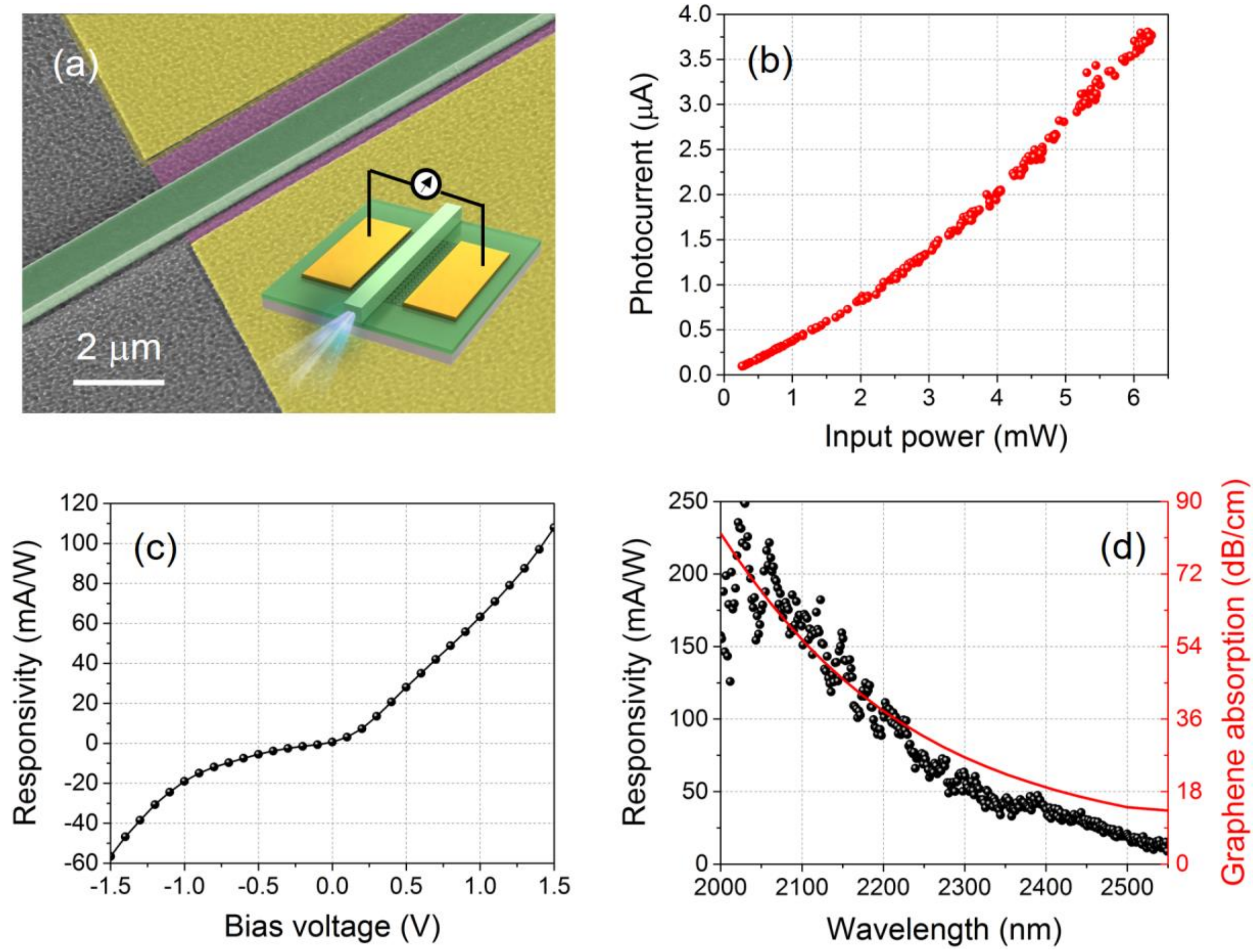

**Fig. 4. (a) SEM tilted-view micrograph of the mid-IR waveguide-integrated detector; inset shows a schematic diagram of the device. (b) Zero-bias photocurrent recorded as a function of input optical power from the waveguide at 2185 nm wavelength. (c) Responsivity of the detector device to 2185 nm waveguide input. (d) Mid-IR broadband spectral dependences of the detector's responsivity (at 1.5 V bias) and calculated optical absorption in the graphene layer.**

Besides simplifying integration of graphene detectors with waveguides on silicon, the use of ChG's further opens up photonic integration on unconventional plastic substrates to enable mechanically flexible photonic systems. Leveraging our previously developed flexible substrate integration protocols[27], we have demonstrated the first waveguide-integrated graphene detector on flexible polymer membranes. Detailed fabrication and characterization outcomes are presented in Supplementary Section XII.

**Broadband mid-IR waveguide modulator**

As previously discussed, the $Ge_{23}Sb_7S_{70}$ glass can function not only as the waveguiding medium, but also as a gate dielectric to control the Fermi level inside graphene. As its Fermi level changes across a threshold value corresponding to half the photon energy, optical absorption of graphene is drastically modified due to Pauli blocking, an effect that has been harnessed to realize near-IR waveguide modulators[3,43-46] and electro-optic manipulation of free-space mid-IR light[47-49]. Here we utilize the versatile ChG material to demonstrate the first graphene-based waveguide modulator operating in the mid-IR. Figure 5a illustrates the device layout and Fig. 5b shows an overlay of the TE modal profile at 2 μm wavelength and an SEM cross-sectional micrograph of the waveguide. The device working principle is similar to that of double-layer graphene modulators developed by Liu *et al.*[50]. In our case, the active region is formed by two graphene sheets separated by a $Ge_{23}Sb_7S_{70}$ glass gate dielectric of 50 nm in thickness. When a gate bias is applied, charges of opposite signs are electrostatically deposited in the two graphene layers, resulting in shifts of their Fermi levels towards opposite directions. Optical transmission in the waveguide (also made of $Ge_{23}Sb_7S_{70}$ glass) is consequently modulated via Pauli blocking. Using this mechanism, we demonstrate broadband optical modulation for the TE mode across the 2.05 μm to 2.45 μm band with modulation depth up to 8 dB/mm as shown in Fig. 5c. A thorough theoretical analysis taking into consideration the starting Fermi levels in the two graphene layers as well as Fermi-Dirac carrier distribution is presented in Supplementary Section XIII. The theoretically predicted waveguide transmittance as a function of gate bias (Fig. 5d) agrees well with experimental measurements. The current device geometry and our characterization setup are not optimized for high-speed tests and limit the modulation time constant to 7 μs, being mainly restricted by the large electrical probe capacitance and series resistance. Our calculations show that with improved device design and measurement schemes the attainable modulation bandwidth can be enhanced by five orders of magnitude to warrant GHz operation using the same device architecture (Supplementary Section XIV).

In summary, we have established a new paradigm for integrating 2-D materials with planar photonic circuits. Unlike traditional methods which rely on post-fabrication transfer, our approach capitalizes on low-temperature ChG deposition to process devices directly on 2-D materials without disrupting their extraordinary optoelectronic properties. In addition to streamlining the 2-D material integration process, our approach envisages novel multilayer structures with unprecedented control of light-matter interactions in the 2-D layers. As an example, we implemented a graphene-sandwiched waveguide architecture to experimentally achieve ultra-broadband on-chip polarization isolation and thermo-optic switching with record energy efficiency. We further leverage the zero-gap nature of graphene to realize ChG waveguide-integrated broadband mid-IR detectors and modulators, the latter of which also makes use of the multifunctional ChG as the gate dielectric for electrostatic tuning of the Fermi level in graphene. We foresee that the versatile glass-on-2D-material platform will significantly expedite and expand integration of 2-D materials to enable new photonic functionalities.

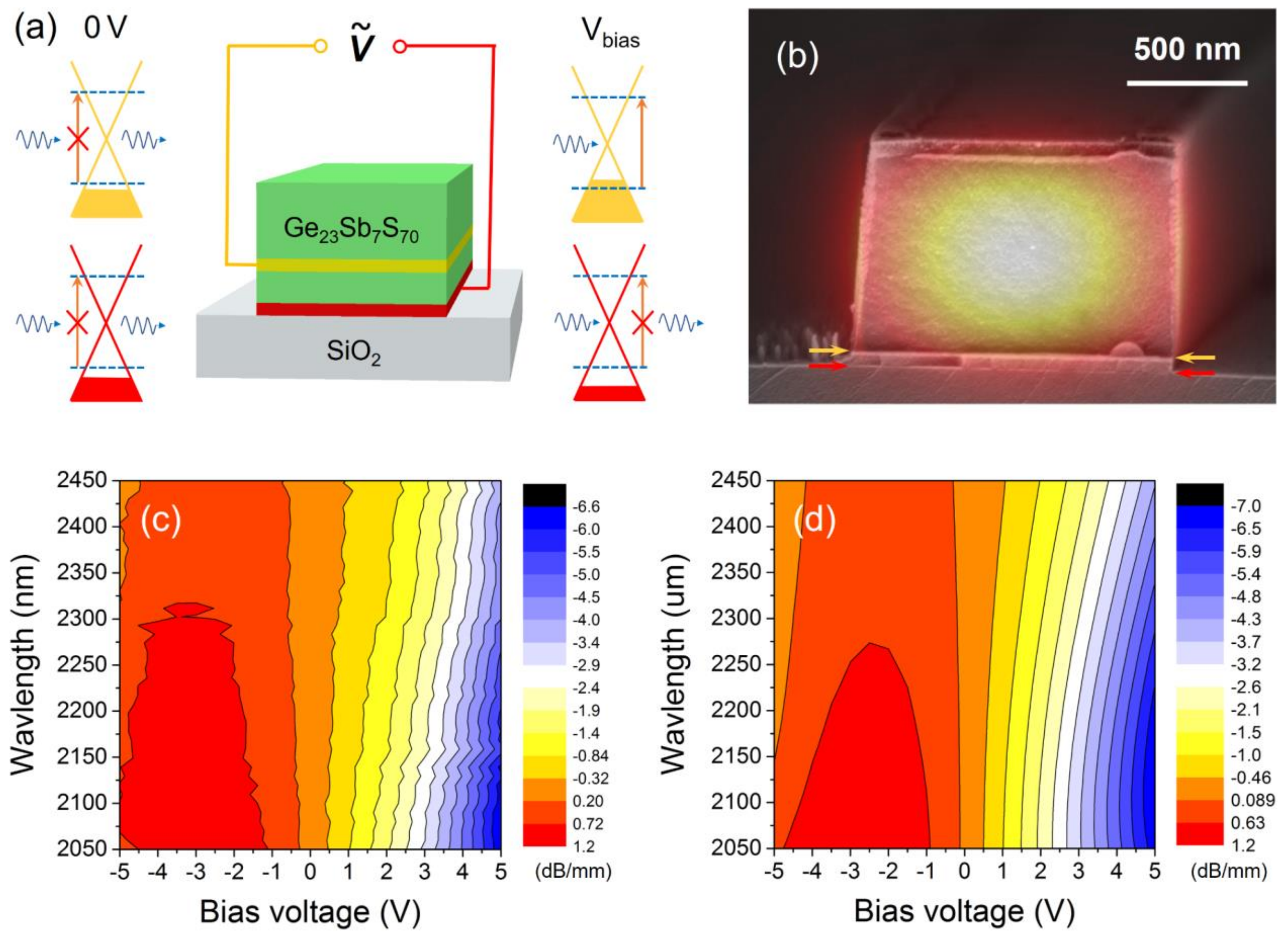

**Fig. 5. (a) Center: schematic diagram of the mid-IR waveguide modulator, where the top and bottom graphene layers are labeled with yellow and red colors, respectively; left and right: band profiles of the two graphene layers, where the brown arrows represent energy of incident photons. (b) Overlay of simulated TE optical mode profile in the modulator waveguide and an SEM cross-sectional image of the device. The arrows point to the locations of the two graphene layers. (c) Measured and (d) simulated color contour maps showing wavelength and bias dependent modulation depth of the device in dB/mm (relative to its transmittance at zero bias).**

## Methods

**Device fabrication.** Device fabrication was performed at the MIT Microsystems Technology Laboratories and the Harvard Center for Nanoscale Systems. For the mid-IR detector and modulator, the starting substrate is a silicon wafer coated with 3 μm thermal oxide (Silicon Quest International), whereas for the polarizer and the thermo-optic switch an additional $Ge_{23}Sb_7S_{70}$ layer was deposited onto the wafer prior to graphene transfer. Monolayer graphene grown using CVD on Cu foils was then transferred onto the substrate following the standard poly(methyl-methacrylate) (PMMA) based wet transfer process[51]. In all cases, the substrate has a planar surface finish, ensuring a high transfer yield. The graphene layer is subsequently patterned using electron beam lithography on an Elionix ELS-F125 electron beam lithography system followed by oxygen plasma etching. Ti/Au (10/50 nm) contact metals were electron beam evaporated and patterned using PMMA as the lift-off resist. A $Ge_{23}Sb_7S_{70}$ glass film is then deposited via thermal evaporation using a custom-designed system (PVD Products, Inc.)[26,52]. Small flakes of $Ge_{23}Sb_7S_{70}$ crushed from bulk glass rods prepared using the standard melt quenching technique were used as the evaporation source material[53]. The deposition rate was monitored in real time using a quartz crystal microbalance and was stabilized at 20 Å/s. The substrate was not actively cooled although the substrate temperature was maintained below 40 °C throughout the deposition as measured by a thermocouple. The $Ge_{23}Sb_7S_{70}$ devices were defined using fluorine-based plasma etching and the detailed etching protocols were discussed elsewhere[54]. If needed, the graphene transfer and glass deposition process can be repeated multiple times to create complex multilayer geometries.

**Device characterization.** The on-chip polarizers were tested using a fiber end-fire coupling scheme and the characterization setup and protocols are described in detail in Supplementary Section VI. The thermo-optic switch devices were measured on a home-built grating coupling system used in conjunction with an external cavity tunable laser (Luna Technologies) with a built-in optical vector analyzer. Laser light was coupled into and out of the devices using single-fiber probes. The DC electrical power was supplied and monitored by a Keithley 2401 Source Measure Unit (SMU). For the dynamic test, the AC electrical power was provided by a Keysight 33521A function generator while the optical output was recorded on an oscilloscope. The mid-IR detector and modulator devices were interrogated using a tunable $Cr^{2+}$:ZnS/Se mid-IR laser covering 2.0 – 2.55 μm wavelengths (IPG Photonics). The mid-IR laser waveguide coupling and real-time wavelength monitoring system is similar to that described in an earlier publication[55] and was illustrated in Supplementary Section XV.

## Acknowledgements

The authors gratefully thank Ren-Jye Shiue and Dirk Englund for helpful discussions on graphene photodetector design, Lionel C. Kimerling and Anu Agarwal for providing access to device measurement facilities, Qingyang Du, Jerome Michon, and Yi Zou for assistance with device processing and characterization, and Mark Mondol for technical support with electron beam lithography. Funding support is provided by the National Science Foundation under award numbers 1453218 and 1506605. This material is based upon work supported by the National Science Foundation Graduate Research Fellowship under Grant No. 1122374. The authors also acknowledge fabrication facility support by the MIT Microsystems Technology Laboratories and the Harvard University Center for Nanoscale Systems, the latter of which is supported by the National Science Foundation under award 0335765.

## Author contributions

H.L. conceived the device designs and carried out device fabrication and testing. Y.S. prepared and characterized the 2-D materials. Y.H. characterized the polarizer and thermo-optic switch devices. D.K. constructed the mid-IR testing system and measured the detector and modulator devices. K.W. performed numerical modeling of the thermo-optic switch. J.L. and Z.H. deposited the ChG films. L.L. and Z.L. contributed to device characterization. S.D.J. performed Raman and passivation tests. S.N. and A.Y. synthesized the ChG materials. H.W. and C.-C. H. assisted with 2-D material preparation. J.H., T.G., J.K.,

K.R., and D.H. supervised and coordinated the research. All authors contributed to technical discussions and writing the paper.

## Competing financial interests

The authors declare no competing financial interests.

**Supplementary Information**

# Chalcogenide Glass-on-Graphene Photonics

**Hongtao Lin[1,*], Yi Song[2], Yizhong Huang[1], Derek Kita[1], Kaiqi Wang[1], Lan Li[1], Junying Li[1,4], Hanyu Zheng[1], Skylar Deckoff-Jones[1], Zhengqian Luo[1,3], Haozhe Wang[2], Spencer Novak[5], Anupama Yadav[5], Chung-Che Huang[6], Tian Gu[1], Daniel Hewak[6], Kathleen Richardson[5], Jing Kong[2], Juejun Hu[1,*]**

[1]*Department of Materials Science & Engineering, Massachusetts Institute of Technology, Cambridge, USA*
[2]*Department of Electrical Engineering & Computer Science, Massachusetts Institute of Technology, Cambridge, USA*
[3]*Department of Electronic Engineering, Xiamen University, Xiamen, China*
[4]*Key Laboratory of Optoelectronic Technology & System, Education Ministry of China, Chongqing University, Chongqing, China*
[5]*The College of Optics & Photonics, University of Central Florida, Orlando, USA*
[6]*Optoelectronics Research Centre, University of Southampton, Southampton, UK*

*hometown@mit.edu, hujuejun@mit.edu

In this Supplementary Information, we provide further details on device simulation, fabrication, and characterization results. This Supplementary Information comprises the following Sections:

## Section I – ChG integration with non-graphene 2-D materials

We have validated ChG integration on four other 2-D materials besides the zero-bandgap graphene: $MoS_2$, black phosphorus (BP), InSe, and hexagonal BN (hBN). The four materials represent four important classes of van der Waals crystals of considerable interest to photonic applications[1]: transition metal dichalcogenide (TMDC), narrow bandgap elemental semiconductor, III-VI layered semiconductor (which exhibit extraordinary optoelectronic properties as well as large second order nonlinearity[2,3]), and insulator.

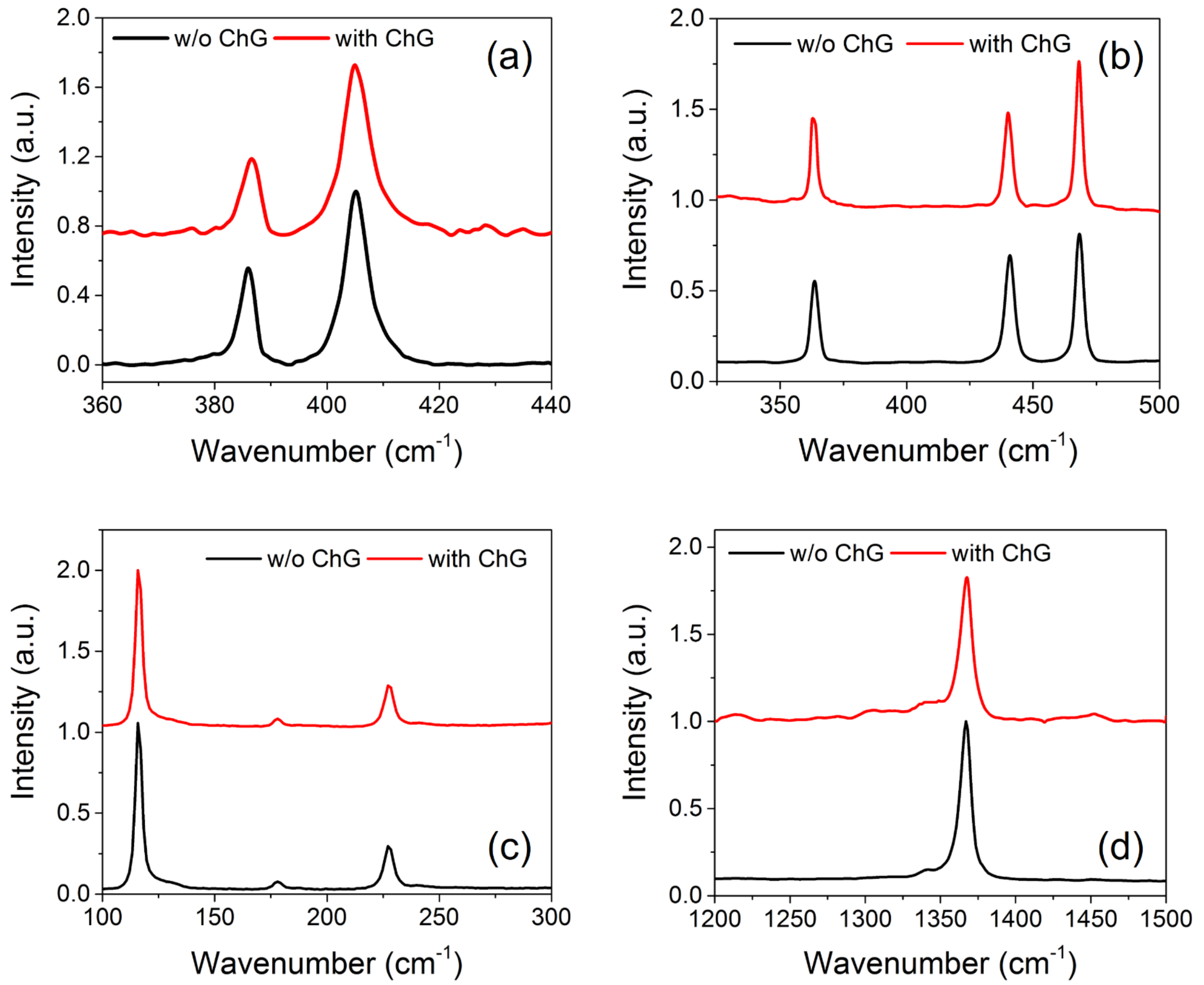

**Fig. S1. Raman spectra of 2-D materials prior to and after deposition of $Ge_{23}Sb_7S_{70}$ glass layer on top: (a) $MoS_2$; (b) black phosphorus; (c) InSe; (d) hBN. Background Raman signals from the $Ge_{23}Sb_7S_{70}$ glass films have been subtracted.**

Monolayer $MoS_2$ was deposited using CVD on an oxidized silicon wafer[4]. We prepared thin sheets of BP (~ 50 nm), InSe (~ 100 nm) and hBN (~ 20 nm) on oxidized silicon wafers by exfoliation from bulk crystals. Details of the exfoliated 2-D crystal sample preparation protocols are described elsewhere[5]. A 50-nm-thick $Ge_{23}Sb_7S_{70}$ ChG film was subsequently deposited onto the 2-D materials using single-source thermal evaporation. Raman spectra of the samples prior to and after ChG deposition were taken on a Raman microscope (LabRAM HR Evolution system, HORIBA Scientific Instruments & Systems) using 532 nm excitation wavelength for $MoS_2$ and 633 nm wavelength for other samples. Figure S1 present the measured Raman spectra. In all cases, the Raman spectra remain unchanged after ChG coating, indicating that the low-temperature ChG film deposition process does not alter atomic structures of the 2-D materials. It is worth noting that background signals from the $Ge_{23}Sb_7S_{70}$ glass films were calibrated by performing Raman measurements on glass films deposited in areas without 2-D material coverage and then subtracted from the raw data to obtain the spectra.

**Section II – ChG passivation of 2-D materials**

In addition to acting as a low-loss light guiding medium and a gate dielectric, ChG's can also play the role of a passivation layer on less chemically stable 2-D materials and effectively prevent their degradation. To illustrate the passivation function of ChG films, uncoated exfoliated black phosphorus flakes (~ 50 nm thick) and flakes coated with 50-nm-thick $Ge_{23}Sb_7S_{70}$ glass were both immersed in 30% $H_2O_2$ solution for 30 s. Figures S2a to S2d compare the morphologies of coated and uncoated BP flakes before and after $H_2O_2$ exposure. It is apparent that the unprotected BP flake was almost completely etched by $H_2O_2$, whereas the coated BP flake exhibited little morphological change. The passivation capability of ChG coating is further evidenced by the Raman spectra taken on BP flakes pre- and post-$H_2O_2$ treatment (Figs. S2e and S2f), showing that the ChG coating preserves the chemical structure of BP in a corrosive environment. In the same vein, our prior work has also demonstrated that a 35-nm-thick $Ge_{23}Sb_7S_{70}$ glass film can effectively prohibit surface oxidation on selenide waveguides and significantly prolong lifetime of the waveguide devices in an ambient environment[6].

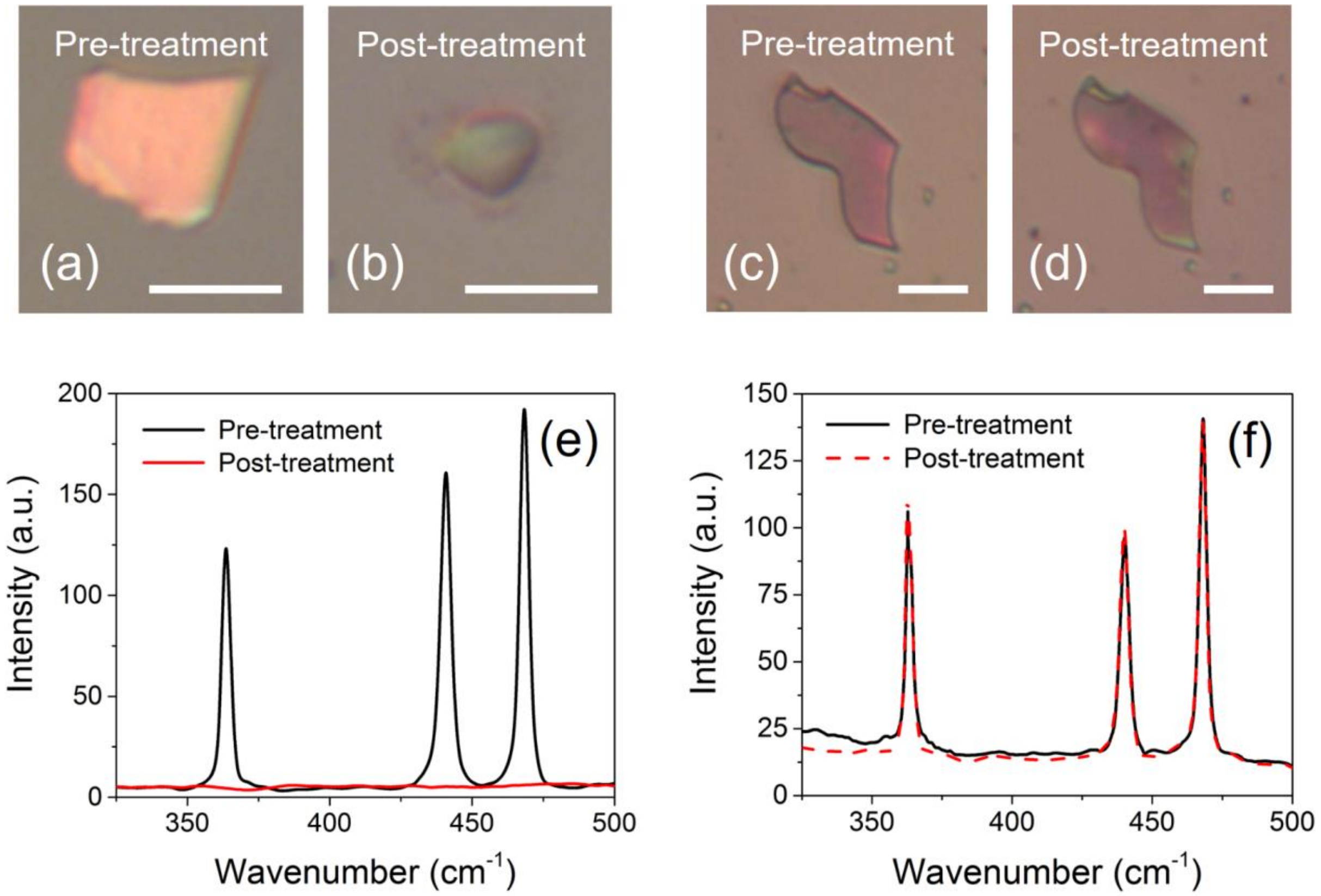

**Fig. S2. (a-d) Optical micrographs on BP flakes (a) without ChG coating and before exposure to $H_2O_2$ solution; (b) without ChG coating and after 30 s immersion in 30% $H_2O_2$ solution; (c) coated with 50 nm $Ge_{23}Sb_7S_{70}$ and before exposure to $H_2O_2$ solution; (d) coated with 50 nm $Ge_{23}Sb_7S_{70}$ and after 30 s immersion in 30% $H_2O_2$ solution. The scale bars are 40 μm in length. (e, f) Raman spectra of BP prior to and after $H_2O_2$ treatment; the spectra were collected on BP flakes (e) without ChG coating; and (f) coated with 50 nm $Ge_{23}Sb_7S_{70}$ glass passivation layer, where background Raman signal from the $Ge_{23}Sb_7S_{70}$ glass coating has been subtracted.**

### Section III – ChG integration on graphene via solution processing and soft nanoimprint

Our paper demonstrated integration of thermally evaporated $Ge_{23}Sb_7S_{70}$ glass with a wide variety of 2-D materials. Here we have shown that the integration process can also make use of other glass compositions and ChG films derived from organic solutions other than vacuum deposition[7-14]. As an example, Fig. S3 schematically illustrates the process of optical grating fabrication in an $As_2Se_3$ glass film deposited on graphene. The glass solution preparation and film coating processes are described in detail elsewhere[11]. Fig. S4a shows an SEM micrograph of a 450-nm-thick $As_2Se_3$ glass film deposited on graphene, which exhibits a smooth, featureless surface. The film has a refractive index of 2.7 at 1550 nm wavelength as determined by ellipsometry measurements. The glass film is subsequently imprinted using a replica molded elastomer stamp following our previously established protocols[10]. Fig. S4b displays an SEM image of a grating with 2 μm period imprinted in the $As_2Se_3$ film showing excellent pattern fidelity. Similar outcomes (not presented here) were also obtained via thermal nanoimprint[15] in evaporated $As_{20}Se_{80}$ films. These results suggest that our facile integration scheme can potentially be implemented using different ChG compositions offering vastly different optical properties (e.g., refractive indices) to fulfill diverse optical device design needs.

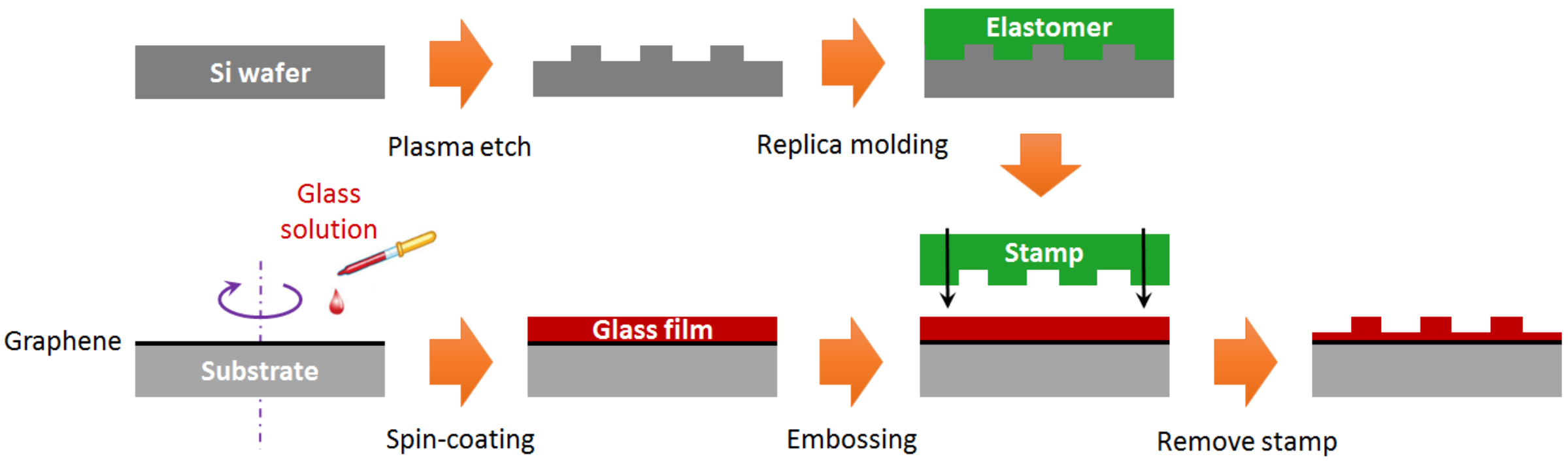

**Fig. S3. Schematic process flow showing integration of solution-processed and nanoimprinted $As_2Se_3$ devices on graphene**

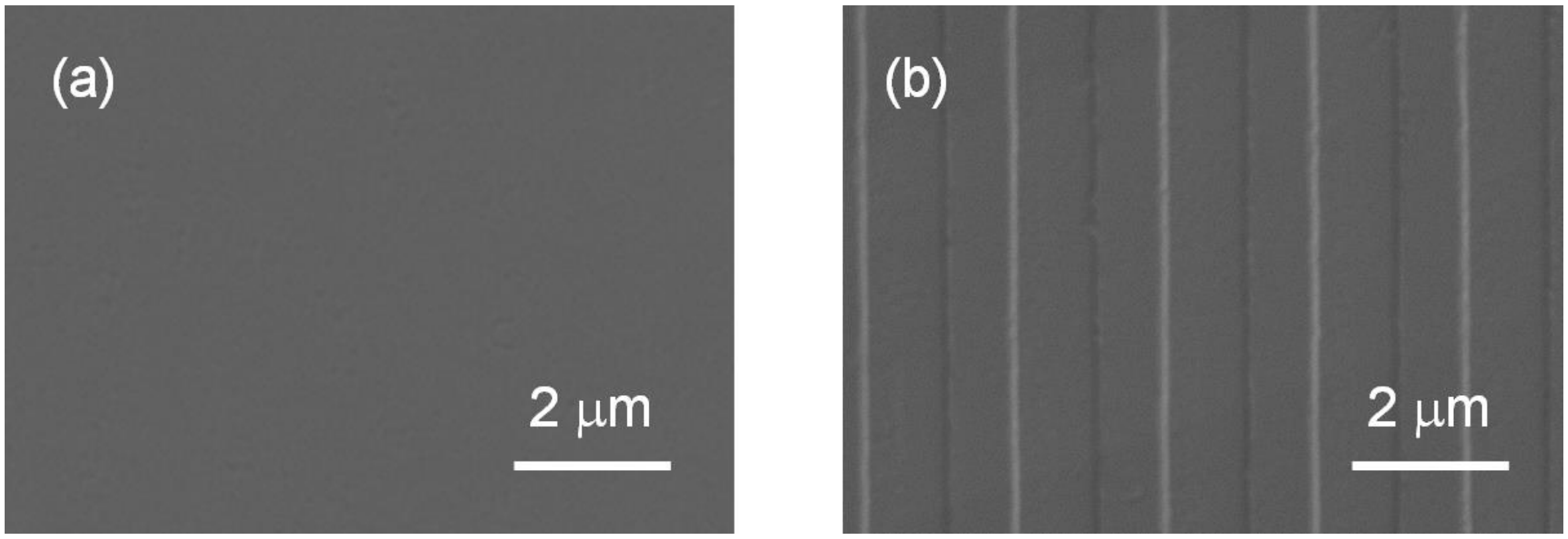

**Fig. S4. (a, b) SEM top-view images of (a) a blanket $As_2Se_3$ film on graphene; and (b) an imprinted grating in $As_2Se_3$ film on graphene.**

### Section IV – Origin of TM mode propagation loss in graphene-sandwiched ChG waveguide

Anti-symmetry of the in-plane electric field components dictates that the TM polarized mode in theory experiences zero optical absorption from the graphene layer. Our experimental measurements however revealed that the graphene layer introduced an excess loss of 20 dB/cm at 1550 nm wavelength for the TM mode. To elucidate the origin of the TM-mode loss, we consider two possible loss mechanisms: 1) thickness deviation of the glass layers; and 2) unevenness of graphene.

When the top and bottom glass layers have different thicknesses, the graphene layer is not located at the node of the in-plane electric fields, resulting in non-vanishing optical absorption of the TM mode. The thickness deviation is schematically illustrated in Fig. S5a. The impact of such non-ideality on optical modal losses was modeled using the finite element method (FEM) implemented through a commercial software package (MODE Solutions, Lumerical Solutions Inc.). Figs. S5b and S5c plot the simulated absorption losses of the TE and TM modes in the graphene-sandwiched waveguide as a function of graphene layer position offset (as defined in Fig. S5a). The color bars indicate the maximum glass film thickness deviation we observed experimentally. It is therefore clear that such thickness deviation is not the main contributor to the measured TM mode optical loss.

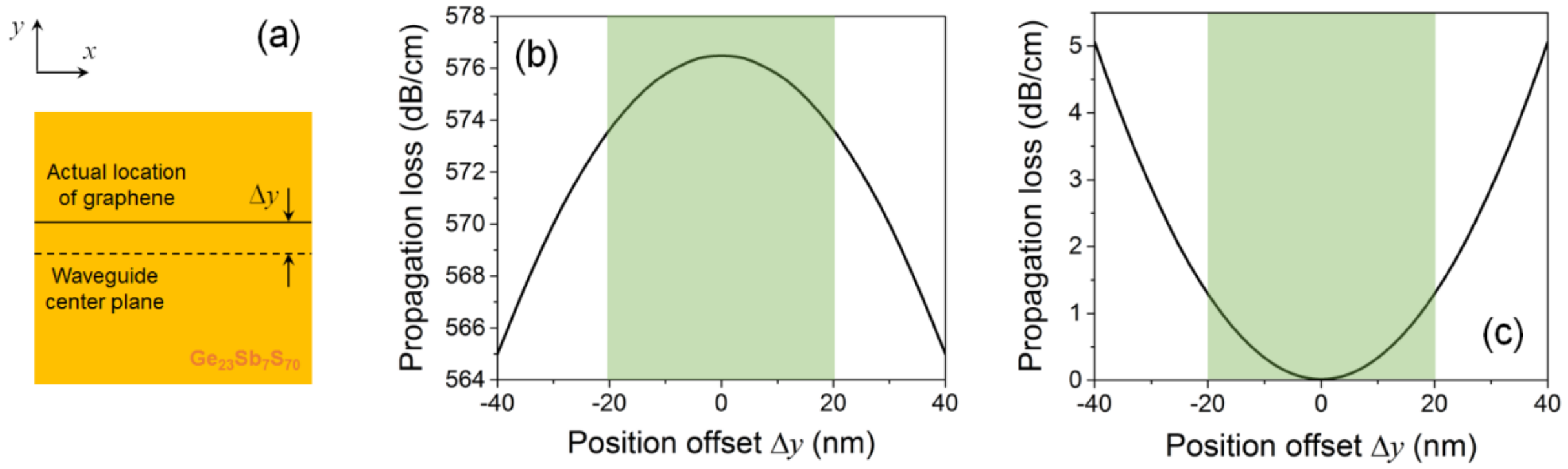

**Fig. S5. (a) Schematic graphene-sandwiched waveguide cross-section showing the misplaced graphene layer due to glass film thickness deviation; (b, c) optical absorption loss due to graphene for the (b) TE and (c) TM modes as functions of the graphene layer location.**

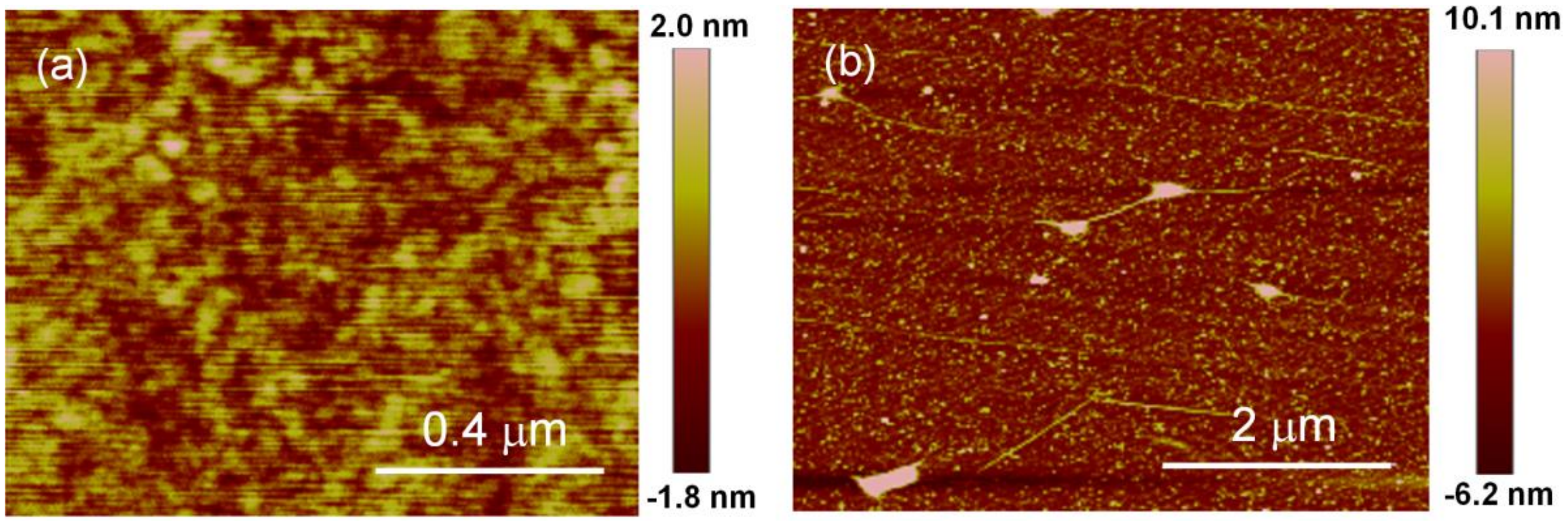

**Fig. S6. Surface morphology of (a) as-deposited $Ge_{23}Sb_7S_{70}$ film; and (b) the same film after graphene transfer.**

To further investigate the origin of the TM mode loss, we measured surface profiles of deposited chalcogenide glass films before and after graphene transfer using atomic force microscopy (AFM), and the results are presented in Fig. S6. The RMS surface roughness increases from 0.5 nm to 2.5 nm after CVD graphene transfer onto the $Ge_{23}Sb_7S_{70}$ film. The increased roughness is attributed to PMMA residue. The PMMA residue can serve as optical scattering centers and increases optical scattering loss. More importantly, it distorts the graphene such that the graphene layer becomes non-parallel to the substrate surface around the PMMA residue. As a consequence, the $E_y$ field component (which reaches maximum at the location of the graphene layer for the TM mode) can also inflict optical absorption loss. Such loss can be mitigated by adopting improved graphene transfer protocols which were shown to effectively eliminate post-transfer polymer residue[16].

## Section V – Polarization-dependent loss measurement of graphene-sandwiched waveguides

Figure S7 shows a schematic diagram of the grating coupler measurement setup used to characterize the micro-ring resonator and Mach-Zehnder interferometer devices, both of which comprise the graphene-sandwiched waveguide. Light from an optical vector analyzer (LUNA OVA 5000) is coupled into an optical fiber and amplified by an erbium doped fiber amplifier when needed. Light from the fiber is coupled into or out of on-chip photonic devices through a pair of grating couplers with an unoptimized coupling loss of approximately 8 dB per coupler. The optical output is sent back to the optical vector analyzer to collect the spectral information.

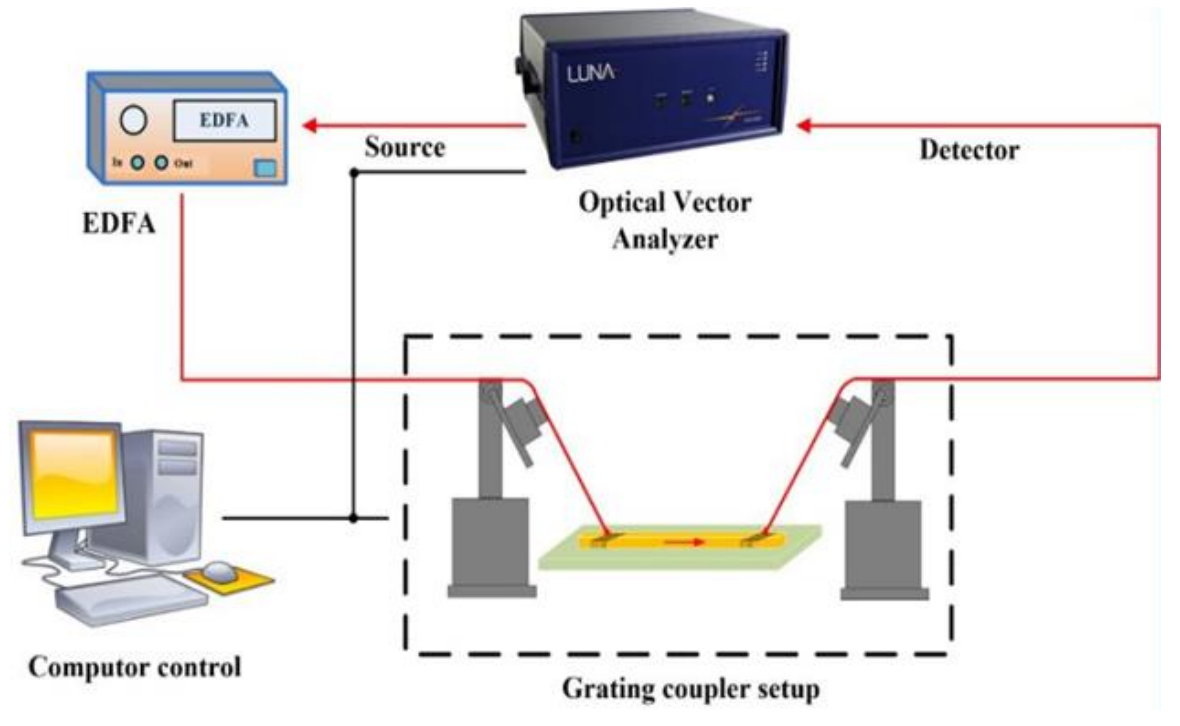

**Fig. S7. Grating coupler measurement system**

We used ring resonator as a highly sensitive vehicle to accurately quantify the relatively low TM mode loss. Figure S8a shows an optical micrograph of a 40-μm-radius ring resonator and Fig. S8b plots the transmission spectrum of the resonator. Propagation loss of the ring resonator is extracted using the coupled-wave transfer matrix method[17]:

$$Q_{load} = \frac{\pi\, L\, n_g}{\lambda_r |\kappa|^2} = \frac{\lambda_r}{\delta\lambda}$$

$$T_{min} = \frac{(\alpha - |t|)^2}{(1 - \alpha|t|)^2} = 10^{-ER/10}$$

$$FSR = \frac{{\lambda_r}^2}{n_g\, L}$$

$$|\kappa|^2 + |t|^2 = 1$$

Here $\kappa$ and $t$ denote the coupling coefficients of the micro-ring coupler, $\delta\lambda$ is the full width at half maximum (FWHM) of the resonant peak in linear scale, $\alpha$ is the propagation loss in the micro-ring, and $L$ represents the round-trip micro-ring length. Other parameters are defined as shown in Fig. S8b.

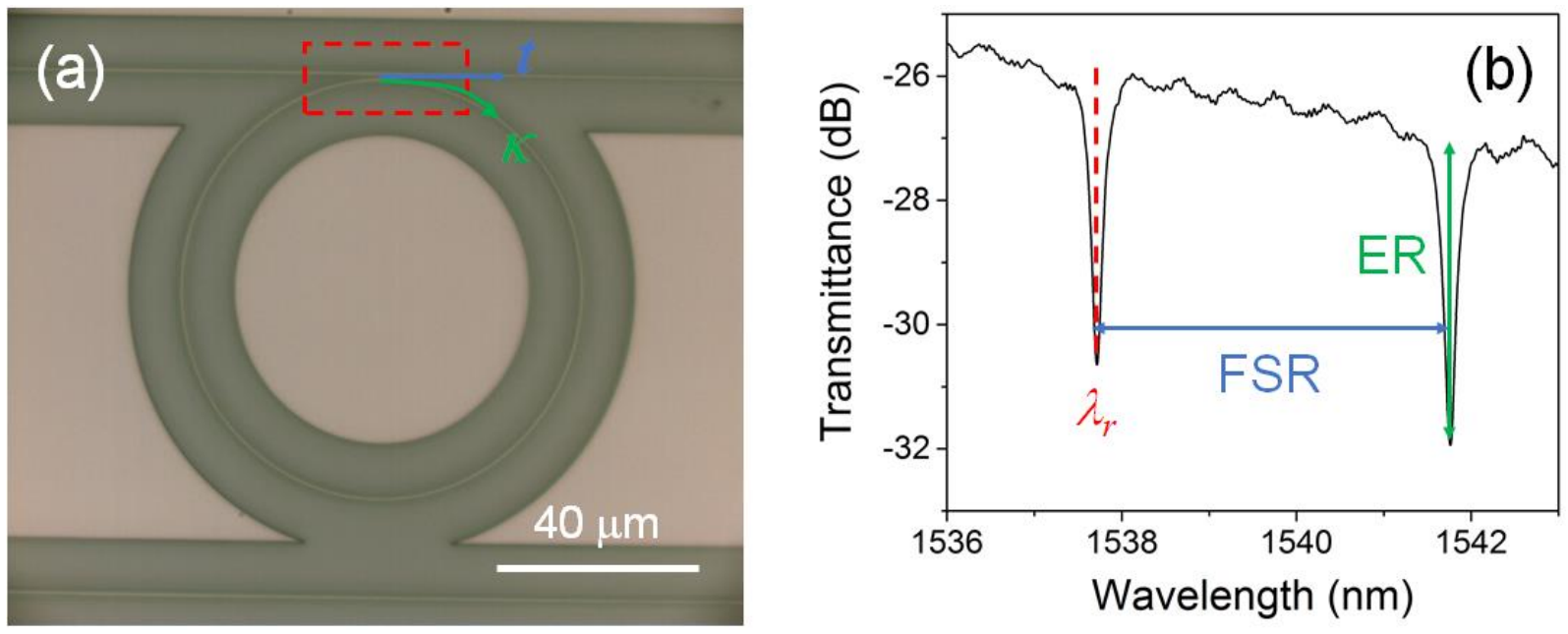

**Fig. S8. (a) Optical micrograph of a graphene-embedded ring resonator; (b) transmission spectrum of the ring resonator.**

Since the graphene-integrated waveguide's TE mode exhibits much higher optical loss, we chose to use unbalanced MZIs for TE mode loss characterization. Compared to the classical cut-back method, the MZI-based approach is inherently immune to coupling variations from device to device due to misalignment or facet damage[18]. Figure S9b presents the transmittance spectra of two MZI devices. Both devices have unbalanced arm lengths (which accounts for the fringes on the transmission spectra), and a graphene patch of varying length is embedded inside each MZI arm. For device 1, its two arms have identical graphene length, whereas for device 2 the graphene embedded sections have different lengths. As a result, the two arms of device 2 experience different optical attenuation, which diminishes the fringe extinction ratio. Optical transmittance $T$ of the MZI is given by:

$$T = \frac{1}{4}[e^{-\alpha l_1} + e^{-\alpha l_2} + e^{-\frac{\alpha(l_1+l_2)}{2}} \cos(\frac{2\pi n_{eff} \Delta L}{\lambda})]$$

where $\Delta L$ is the MZI arm length difference, and $l_1$ and $l_2$ are the graphene embedded section lengths in two MZI arms. The equation yields extinction ratio (ER) of the MZI, i.e. the ratio of the maximum transmittance over the minimum transmittance, as:

$$ER = \frac{T_{max}}{T_{min}} = \frac{e^{-\alpha l_1} + e^{-\alpha l_2} + 2e^{-\frac{\alpha(l_1+l_2)}{2}}}{e^{-\alpha l_1} + e^{-\alpha l_2} - 2e^{-\frac{\alpha(l_1+l_2)}{2}}} = (\frac{1 + e^{-\frac{\alpha \Delta l}{2}}}{1 - e^{-\frac{\alpha \Delta l}{2}}})^2 \approx (\frac{4}{\alpha \Delta l})^2$$

where $\Delta L$ is the length difference for the unbalanced MZI, and $\Delta l$ denotes the graphene embedded section length difference between the two arms. Optical absorption induced by graphene can therefore be inferred from the ER.

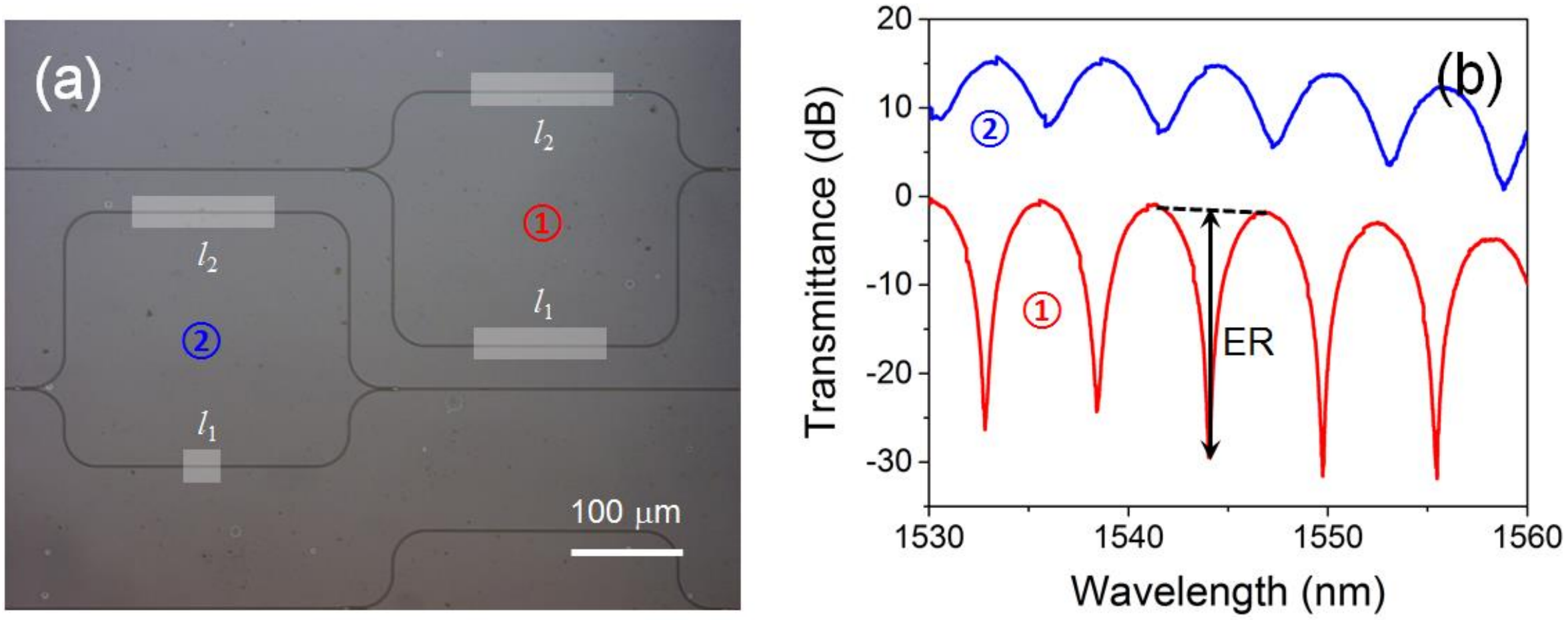

**Fig. S9. (a) Optical micrograph of two unbalanced MZI devices used for TE mode loss characterization: the gray boxes on the MZI arms label the waveguide sections within which a graphene patch is embedded; (b) transmission spectra of the two MZI devices. TE mode optical loss of the graphene sandwiched waveguide can be calculated from the device extinction ratio.**

**Section VI – Broadband operation of graphene polarizer: characterization, analysis, and performance benchmark**

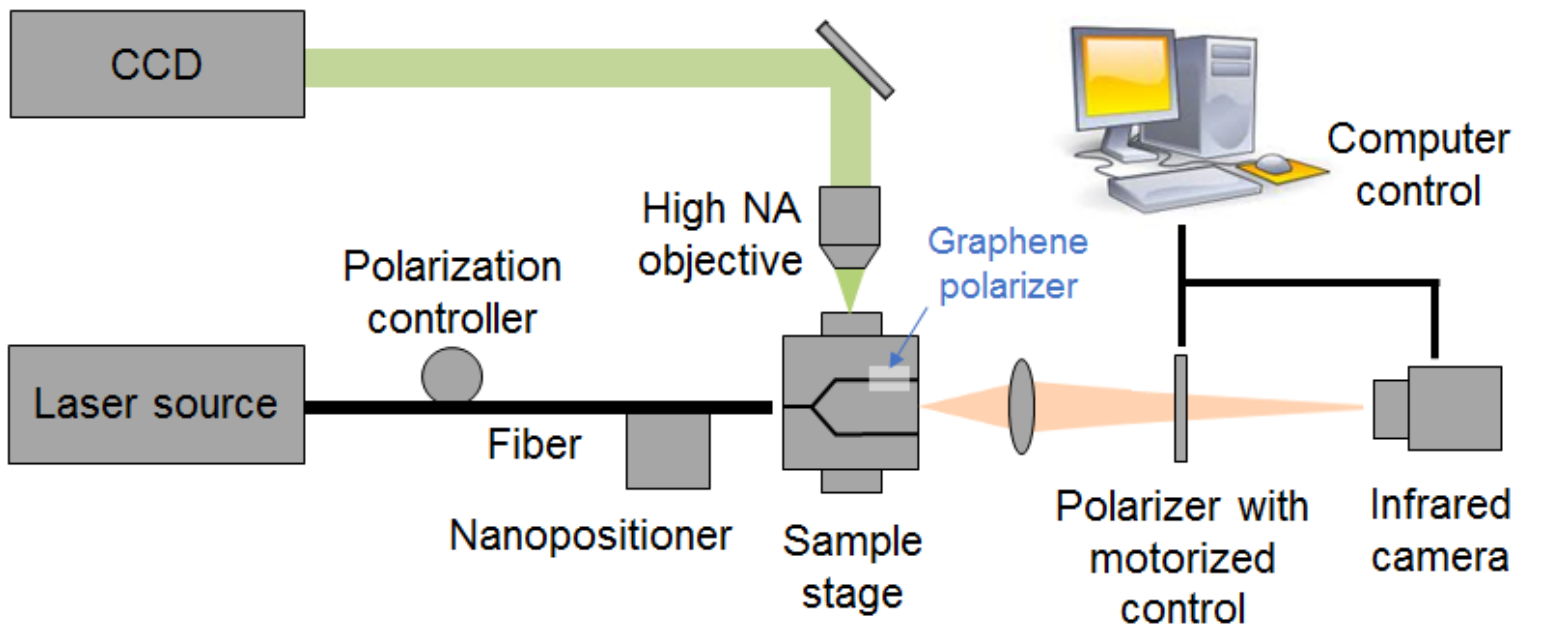

**Fig. S10. On-chip graphene polarizer measurement setup**

Fig. S10 shows a schematic diagram of the measurement setup used to characterize the on-chip graphene waveguide polarizer. Two lasers were employed in the test, a 1550 nm external cavity tunable laser (Luna Technologies) and a 980 nm butterfly laser diode. The laser light was coupled into the on-chip device via a tapered lens-tip fiber probe. The laser light is elliptically polarized when it exits from the fiber probe. A 3-dB waveguide splitter divides the optical signal into two waveguide arms, one of which contains an embedded graphene layer while the other serves as a reference. Output from both arms is filtered by a free-space polarizer and then imaged in the far field using an infrared camera. Figure S11 present exemplary far field output images when the free-space polarizer is rotated to three different angles (measured with respect to the substrate surface plane). The output optical power from a waveguide was calculated by integrating the signal counts from pixels within the waveguide's mode image area. Output from the reference arm is used to calibrate the light intensity propagating inside the waveguide device at any given polarization angle. The polarizer transmittances at 980 nm and 1550 nm (Fig. 2h) were obtained by normalizing the output power from the polarizer arm to that from the reference arm. We note that there is noticeable stray light from the reference arm overlapping with the output mode from the polarizer arm. While this effect has a negligible impact on the polar plot measurement (Fig. 2h) at polarization angles larger than 0° (i.e. when output power from the polarizer arm is significant), it results in underestimated polarization extinction ratios from our experiment. Indeed, extinction ratio inferred using this technique (ER = 21 dB at 1550 nm wavelength) is slightly lower than that obtained from direct waveguide loss measurement at the same wavelength (ER = 23 dB).

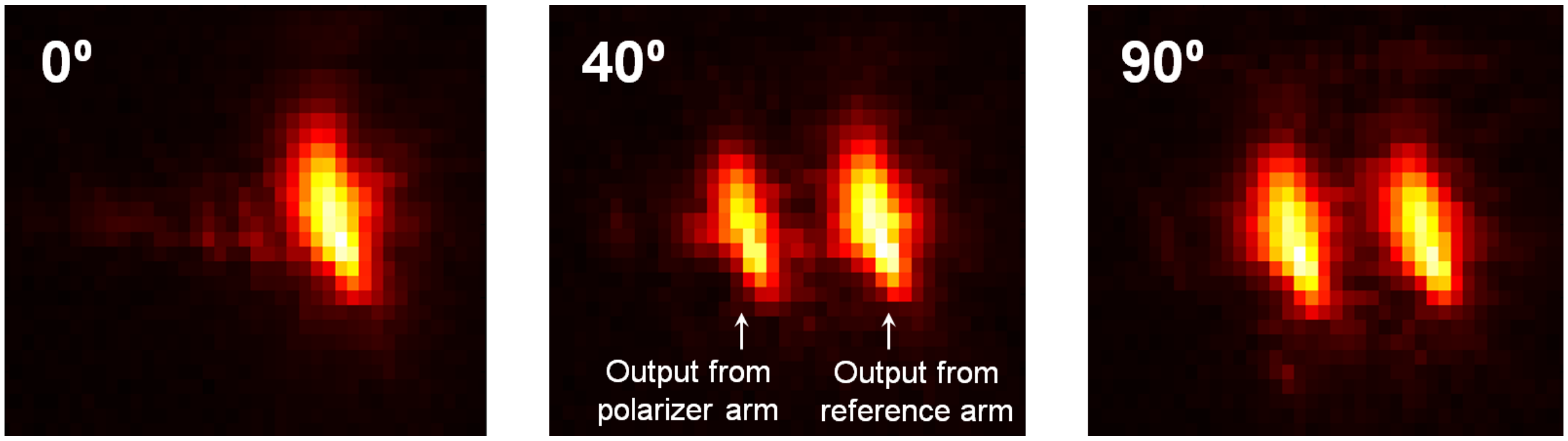

**Fig. S11. Far-field output optical image from the device at 980 nm recorded at different polarizer angles (measured with respect to the substrate surface plane)**

The broadband performance characteristics of the polarizer device can be analyzed by examining the optical modes supported in the graphene-sandwiched waveguide. Fig. S12 plots the modal effective indices of the graphene-sandwiched waveguide as a function of wavelength simulated using FEM. The symmetric modes are marked with solid lines whereas the modes exhibiting anti-symmetry are represented by the dotted lines. The anti-symmetric modes are not excited in our experiments since the input fiber mode is symmetric. It is interesting to note that the polarizer device is multimode at both 1550 nm and 980 nm. To address the polarization rejection mechanism in the multi-mode regime, Fig. S13 depicts all the guided optical modes supported in the graphene-sandwiched waveguide at 980 nm wavelength. For the symmetric TM modes (modes 2 and 7) that can be excited in the polarizer device, all the in-plane electric field components vanish at the center plane of the waveguide where the graphene layer is located, which accounts for the vanishing graphene absorption. On the other hand, in-plane electric field components of the symmetric TE modes (modes 1 and 8) reach maximum at the center plane, leading to large graphene absorption. The first symmetric TM mode with non-vanishing in-plane electric field components is mode 10, and thus its onset at 940 nm defines the lower wavelength bound for our polarizer device operation.

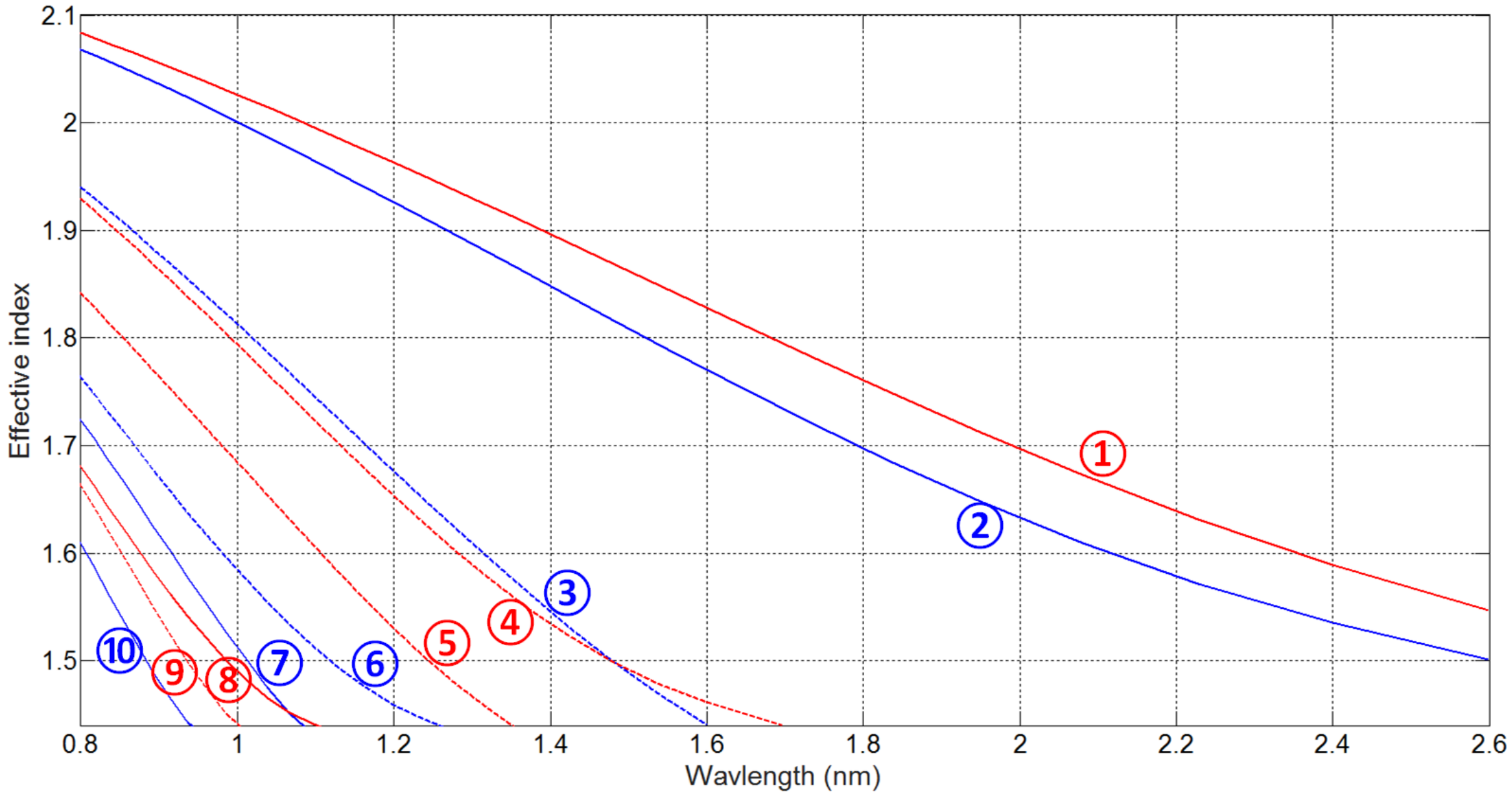

**Fig. S12. Dispersion relation of guided modes in the graphene-sandwiched waveguide modeled using FEM: the solid lines represent symmetric modes whereas the dotted lines label the anti-symmetric modes which are suppressed in our experiment.**

The explanation above is also consistent with our modal loss simulations shown in Fig. S14a for a graphene Fermi level of 0.39 eV below the Dirac point (as is the case in our fabricated polarizer device), and in Fig. S14b when the graphene Fermi level coincides with the Dirac point. We made no intentional effort to engineer the graphene Fermi level during polarizer fabrication; however, the latter case (undoped graphene) can be experimentally realized, for instance, through sandwiching graphene between a pair of hexagonal BN layers[19] or compensating n-doping using ethylene amines[20]. Alternatively, we have demonstrated graphene Fermi level tuning using a chalcogenide gate. The color shadings in the figures label the operation wavelength regime where a 400-μm-long polarizer device exhibits a polarization extinction ratio above 20 dB. When undoped graphene is used, our polarizer design offers ultra-broadband operation spanning over one octave from 0.94 μm to 2.5 μm wavelength.

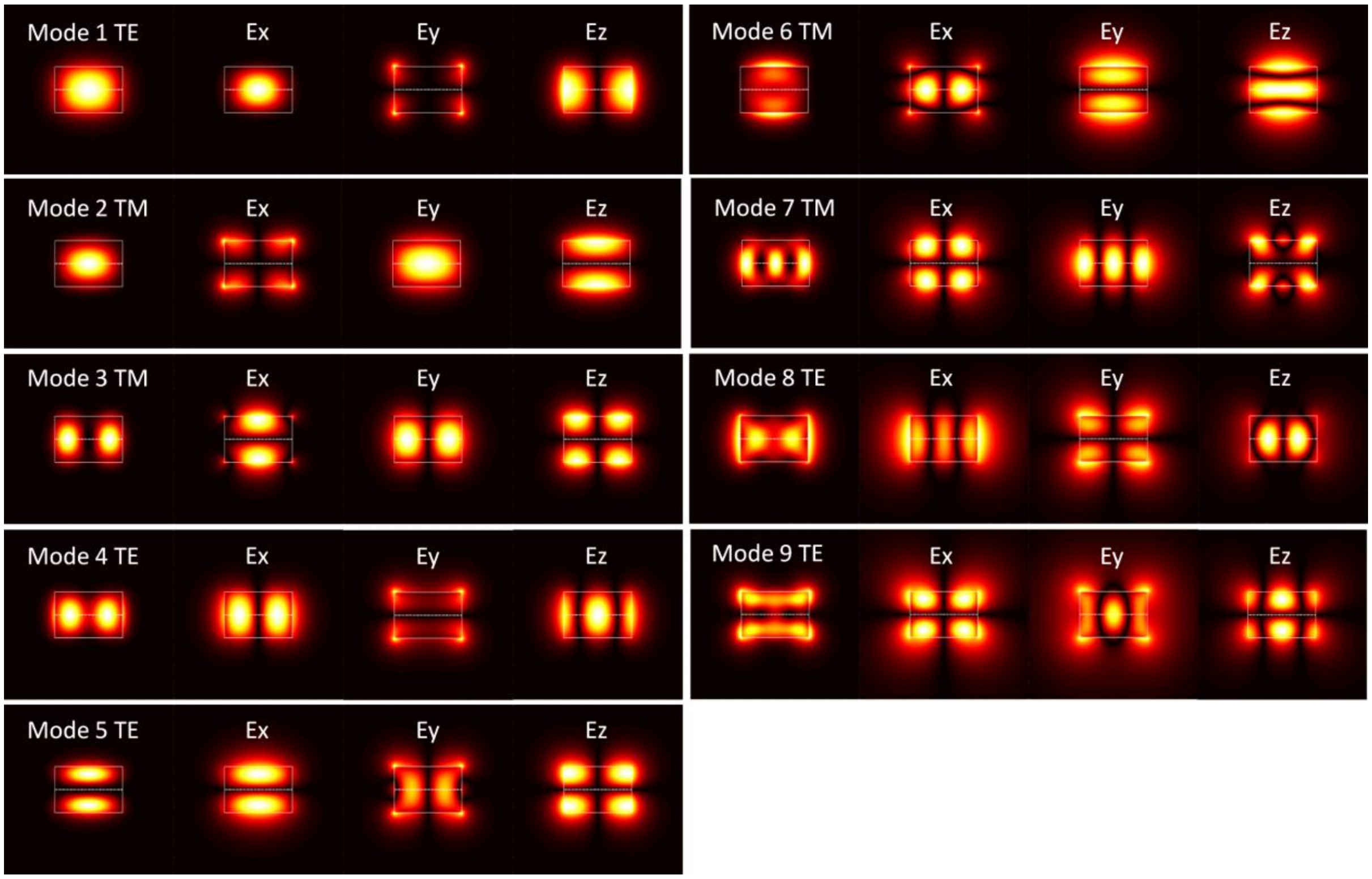

**Fig. S13. Guided modes supported by the graphene-sandwiched waveguide at 980 nm wavelength: the four images (from left to right) for each mode illustrate the distributions of the total electric field intensity, $E_x$, $E_y$, and $E_z$, respectively.**

To benchmark our polarizer performance, Table S1 compares the key metrics of our polarizer device with other on-chip broadband optical polarizers. The performance characteristics of fiber-based graphene polarizers are also included for comparison. Notably, our device claims the lowest insertion loss as well as the widest operation bandwidth among experimentally demonstrated on-chip polarizers.

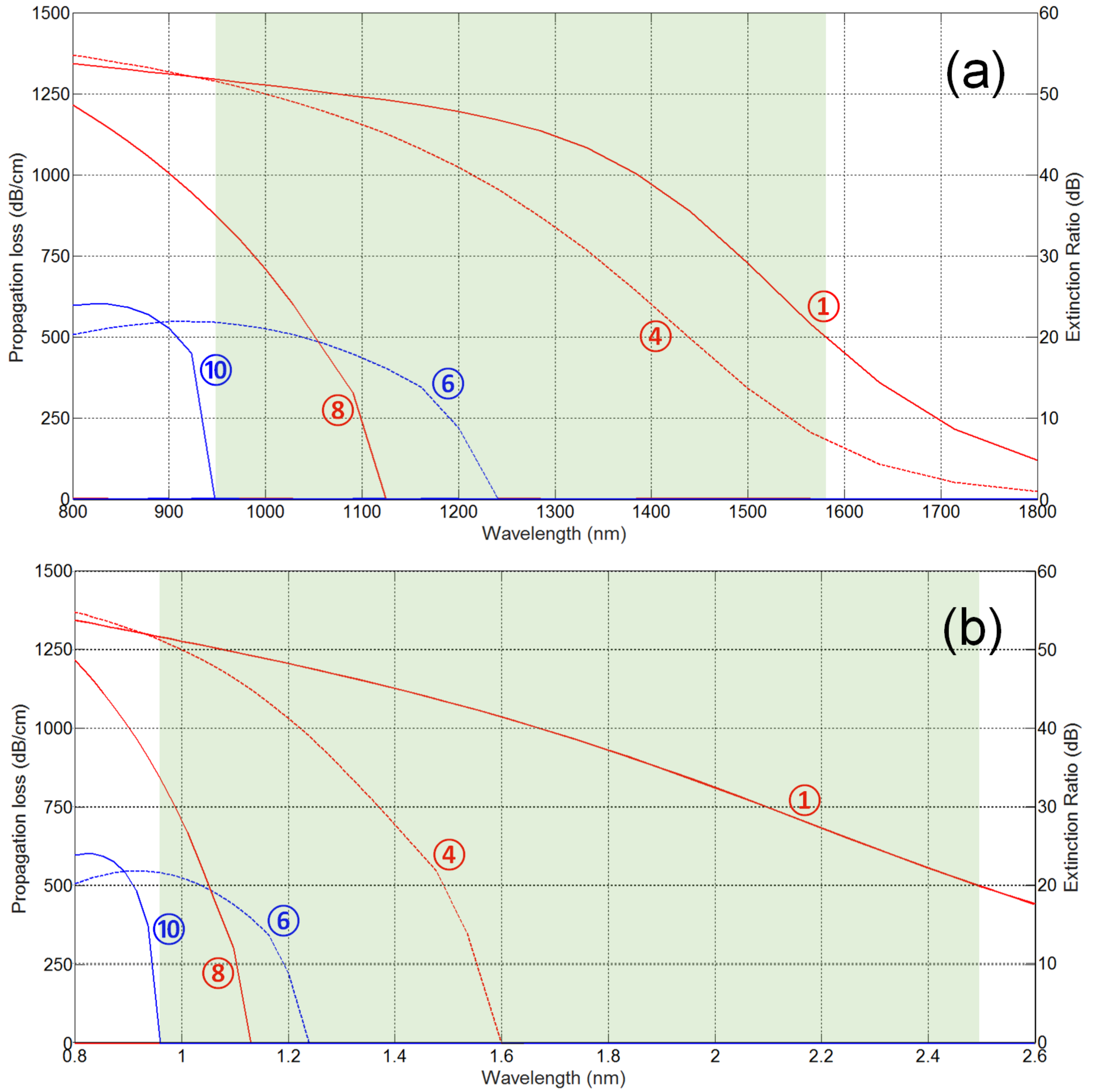

**Fig. S14. Simulated modal propagation losses due to graphene absorption: the graphene Fermi level is set to be (a) 0.39 eV below the Dirac point as is the case in our fabricated polarizer device; and (b) coinciding with the Dirac point. The color shadings mark the wavelength window where a 400-μm-long polarizer device exhibit over 20 dB extinction ratio. The short wavelength end of the operation window is defined by onset of the first symmetric TM-polarized mode (mode 10) at 940 nm. The long wavelength bound of the window is limited by reduced modal overlap and hence decreased absorption in graphene.**

**Table S1. Performance comparison of on-chip polarizers and graphene-based optical polarizers: numbers in black indicate experimentally measured data while numbers in red are theoretical projections; cells with yellow shading represent on-chip polarizer devices. "N/A" indicates that the device is only characterized or modeled at a single wavelength. The FOM is defined by taking the ratio of extinction ratio to insertion loss. SOI: silicon-on-insulator.**

| Device | Length (mm) | Extinction ratio (dB) | Insertion loss (dB) | Figure of Merit (FOM) | Fractional bandwidth (≥ 20 dB ER) |
|---|---|---|---|---|---|
| Graphene-embedded ChG waveguide (this report) | 0.4 | 23 | 0.8 | 29 | 0.45<br>0.91 |
| Metal-dielectric composite loaded waveguide[21] | 0.5 | 20 | ~ 5 | 4.0 | 0.32 |
| Waveguide metal gratings[22] | 1 | 50 | 2 | 25 | 0.14 |
| Shallowly etched waveguide[23] | 1 | 25 | ~ 3 | 8.3 | 0.06 |
| Birefringent polymer cladded waveguide[24] | 2 | 39 | 4 | 9.8 | N/A |
| Horizontal plasmonic slot waveguide[25] | 0.001 | 16 | 2.2 | 7.3 | ER < 20 dB |
| SOI waveguide with modal cut-off[26] | 0.0025 | ~ 25 | ~ 1 | 25 | 0.15 |
| Multilayer hybrid plasmonic waveguide[27] | 0.017 | 30 | ~ 1 | 30 | 0.13 |
| Spiral SiN waveguide[28] | 1,000 | 75 | 2.6 | 29 | 0.08 |
| Subwavelength grating waveguide[29] | 9 | 27 | 0.5 | 54 | 0.039 |
| Graphene-loaded silica waveguide[30] | 4 | 27 | 9 | 3 | 0.24 |
| Graphene-loaded polymer waveguide[31] | 7 | 19 | 26 | 0.73 | N/A |
| Graphene loaded waveguide[32] | 1 | 48 | 10.5 | 4.6 | N/A |
| Graphene-loaded side-polished fiber[33] | 2.1 | 27 | 5 | 5.4 | 1.07 |
| Graphene-loaded side-polished fiber[34] | 5 | 29 | ~ 3 | 9 | 0.12 |
| Graphene-coated surface-core fiber[35] | 3 | 26 | 1.1 | 24 | 0.17 |

## Section VII – Photonic crystal thermo-optic switch device layout

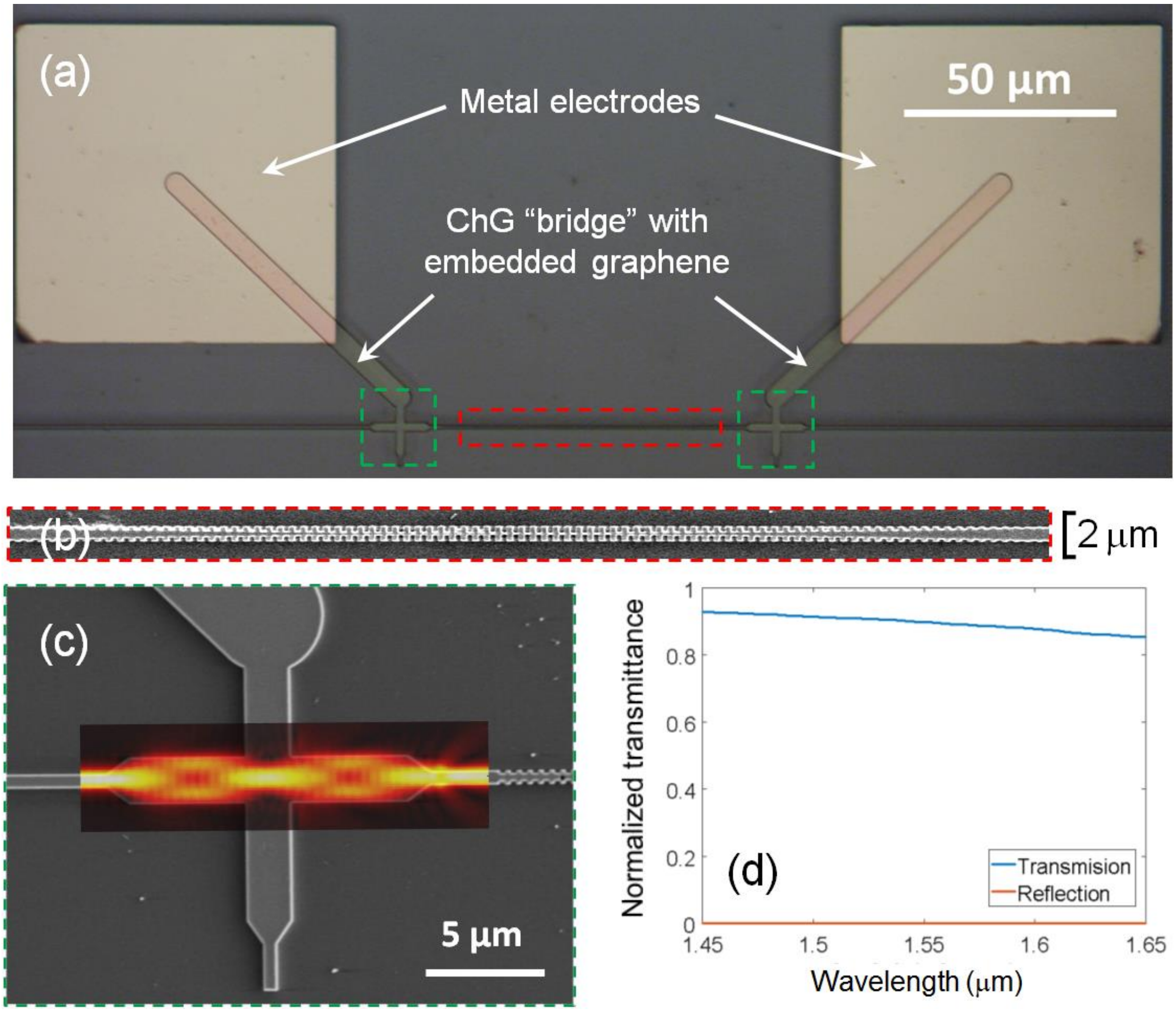

**Fig. S15. (a) Top-view of the fabricated thermo-optic switch device: the embedded graphene layer is connected to two metal electrodes via two ChG "bridges" with embedded graphene layers. The "bridges" have a sandwiched glass-graphene-glass multilayer configuration identical to that of the waveguide photonic crystal. The green and red boxes label the bridge-waveguide crossings and the photonic crystal cavity, respectively; (b) top-view SEM image of the waveguide photonic crystal cavity with embedded graphene; (c) top-view SEM micrograph of a crossing connection between the ChG "bridge" and the waveguide photonic crystal; an image of the simulated optical field distribution inside the structure is superimposed on the micrograph; (d) simulated optical transmittance through the waveguide-bridge crossing showing broadband efficient transmission.**

Figure S15a shows a top-view optical micrograph of the fabricated thermo-optic switch device. The device was fabricated on an oxide-coated silicon wafer. A $Ge_{23}Sb_7S_{70}$ glass film of 280 nm thickness was first deposited via thermal evaporation, followed by graphene transfer to cover the entire substrate. Ti/Au metal electrodes were then deposited and patterned via lift-off on the graphene layer. A second layer of $Ge_{23}Sb_7S_{70}$ glass with an identical thickness of 280 nm was subsequently deposited to encapsulate the graphene layer, which also placed it at the center of the waveguide. Next, the graphene-sandwiched glass layers were patterned via plasma etching to form the waveguide as well as the "bridge" structures between the waveguide and the metal electrodes. The graphene embedded inside the "bridge" and the waveguide was therefore connected to both electrodes, forming a continuous path for current flow.

The bridge-waveguide crossings are engineered to minimize optical loss of waveguide mode propagating through the structures following a multi-mode interferometer (MMI) waveguide crossing design[36]. At the crossing, the waveguide assumes the form of an MMI with a length twice of its self-imaging length[37]. Consequently, an image of the input waveguide mode is formed at the center of the MMI where the MMI intersects with the ChG "bridge". Since the mode image is tightly confined in the center of the MMI as is evident from the optical field distribution in Fig. S15c, scattering loss from the abrupt MMI-bridge junctions as well as optical leakage into the "bridges" are suppressed. Figure S15d plots TM mode optical transmittance through the crossing structure modeled using the finite-difference time-domain (FDTD) technique. The simulation result indicates that the structure exhibits low insertion loss (< 0.7 dB) across the broad spectral range from 1.45 μm to 1.65 μm.

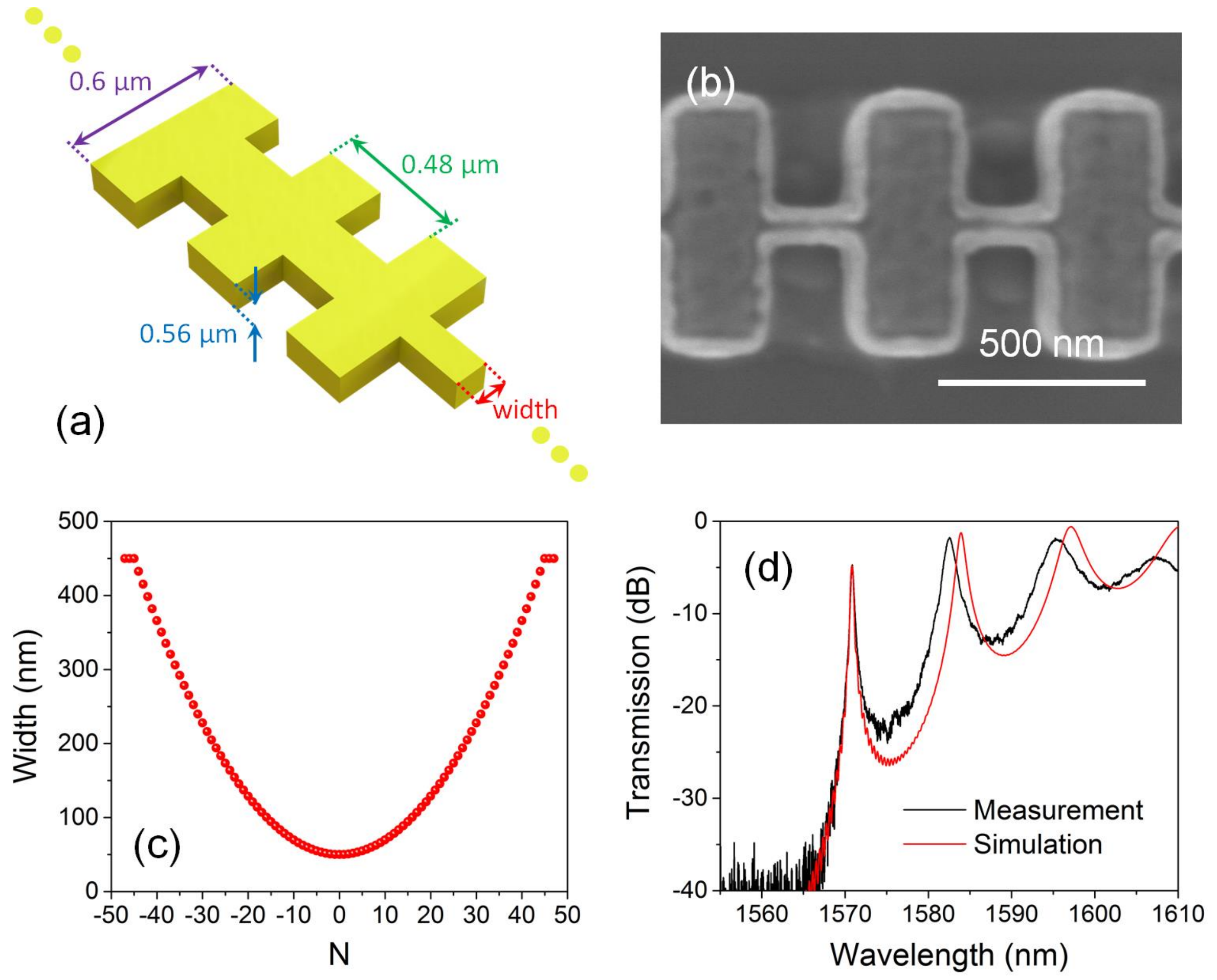

**Fig. S16. (a) Schematic design and (b) SEM top-view image for the photonic crystal switch: the width is modulated along the photonic crystal to create optical confinement; (c) width modulation along the photonic crystal, where N labels number of the repeating unit; (d) normalized transmittance of the photonic crystal cavity: the simulated spectrum is redshifted by 7.2 nm.**

The photonic crystal cavity assumes a width modulated design and the design parameters are listed in Figs. S16a and S16c[38,39]. Figures S15b and S16b show SEM top-view images of the fabricated photonic crystal cavity. The simulated and measured TM mode transmission spectra of the cavity are plotted in Fig. S16d. We note that the simulated spectrum is redshifted by 7.2 nm in the figure. This minor resonant wavelength deviation can be attributed to an index variation ($\Delta n \sim 0.005$) of the glass material from the values used in our design. Other than the slight peak redshift, the experimental data agree well with the simulation results. We measured Q-factors up to 8,000 for the cavity resonant peak near 1570 nm. This Q-factor is considerably higher than values previously reported in graphene-loaded resonators[40,41], despite that in our device graphene is embedded throughout the entire cavity length. The high Q-factor, which is derived from our unique low-loss graphene-sandwiched waveguide design, underlies the superior performance of our thermo-optic switch.

## Section VIII – Coupled thermal/optical modeling of photonic crystal thermo-optic switch

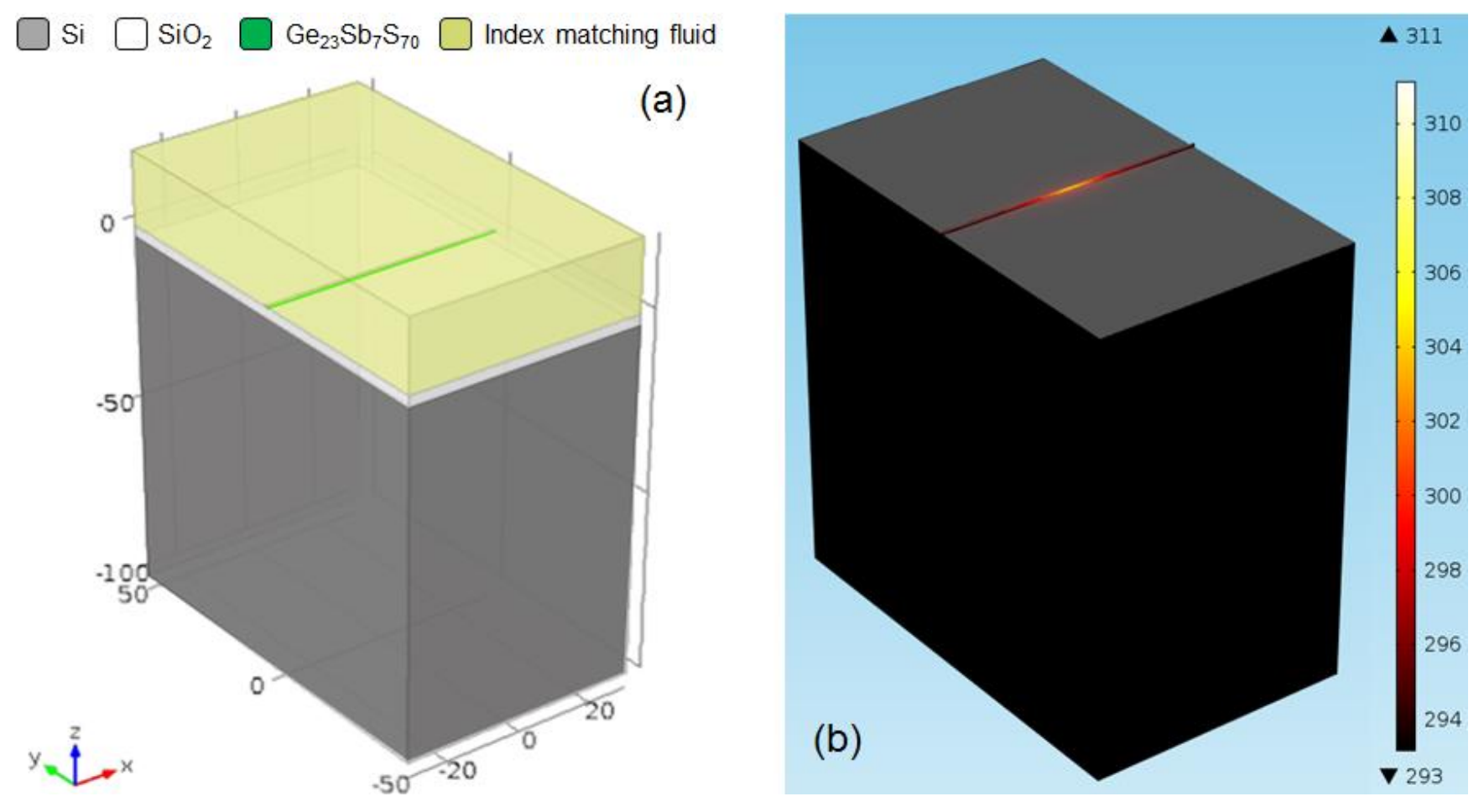

**Fig. S17. (a) Simulation domain in COMSOL; (b) simulated thermal profile of the switch.**

Optical modeling of the photonic crystal cavity switch was performed using the commercial FDTD software FDTD Solutions (Lumerical Solutions Inc.). Finite element method (FEM) thermal simulations were carried out using the COMSOL Multiphysics package. Figure S17a illustrates the thermal simulation domain configuration. At all solid boundaries ($SiO_2$ under cladding and Si substrate), the temperature was fixed at 298K. For the liquid boundaries, a convective heat flux boundary condition was implemented. A 2-D heat source was deduced from electric current spatial distribution and Ohm's law, where the graphene sheet resistance was taken as 400 ohm/sq. based on measured data. We then combined the simulated temperature distribution from COMSOL (Fig. S17b) and modal profile from FDTD modeling to calculate the thermo-optic resonant wavelength shift $\Delta\lambda$ following the classical cavity perturbation formalism[42]:

$$\Delta\lambda = \lambda \cdot \frac{\int (dn/dT)\cdot n(\boldsymbol{r})\Delta T(\boldsymbol{r})\left|E(\boldsymbol{r})\right|^2 dV}{\int n^2(\boldsymbol{r})\left|E(\boldsymbol{r})\right|^2 dV},$$

where $\boldsymbol{r}$ symbolizes the position vector, $\lambda$ corresponds to the unperturbed resonant wavelength, $dn/dT$ denotes the thermo-optic coefficient, $n$ gives the refractive index distribution, $\Delta T$ represents the temperature rise, $E$ is the modal electric field profile, and the integration is performed over the entire simulation domain. The following set of parameters were used to obtain the simulation results presented in Fig. 3: $\lambda$ = 1566 nm, $n_{Ge23Sb7S70}$ = 2.17, $n_{SiO2} = n_{fluid}$ = 1.46, $n_{Si}$ = 3.46, $(dn/dT)_{Ge23Sb7S70} = 2.5 \times 10^{-5}$ /K, $(dn/dT)_{Si} = 1.8 \times 10^{-4}$ /K, $(dn/dT)_{fluid} = -6 \times 10^{-4}$ /K, and $(dn/dT)_{SiO2} = 10^{-5}$ /K. Excellent agreement between our simulations and the experimental results is evident from Fig. 3f.

## Section IX – Lumped element circuit model of thermo-optic switches

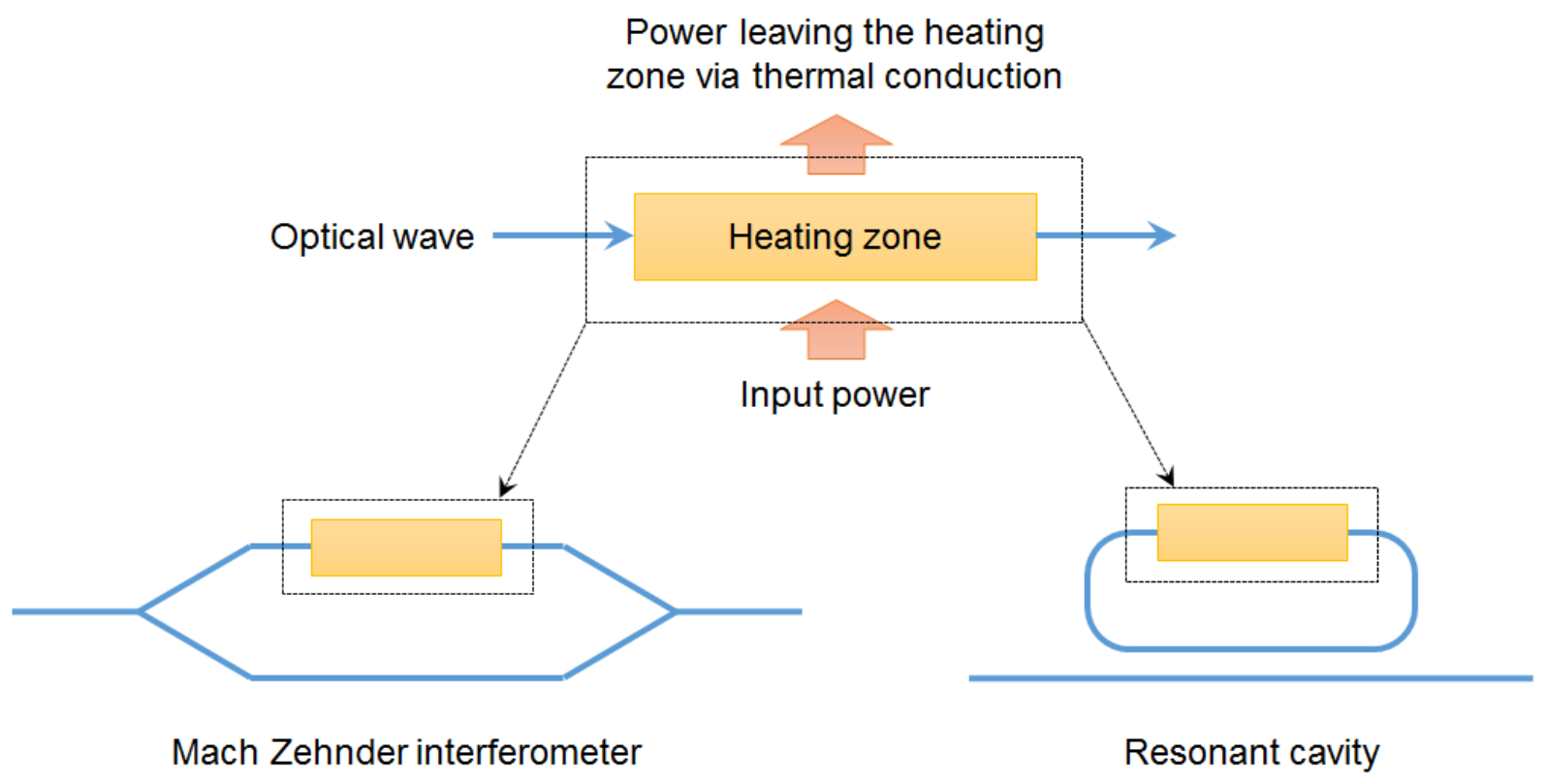

**Fig. S18. Block diagram illustrating the switch configurations**

Here we develop a simple analytical model of thermo-optic switches using lumped element circuits. We consider two switch device configurations, a Mach Zehnder interferometer (MZI) and a resonant cavity. The latter can in principle assume any specific resonant device form including micro-rings, micro-disk, photonic crystal nanobeam cavities, etc. Figure S18 illustrates the two basic device configurations, each including a heating element. Lumped heat capacity of the heating zone is labeled as $C$, and the effective thermal conductance from the heating zone to the heat sink (environment) is denoted as $G$. Thermal time constant $\tau$ of the switch is therefore:

$$\tau = \frac{C}{G}. \tag{X.1}$$

Next let's consider power consumption of the switch in its on-state. Firstly, we specify a thermo-optic phase shift $\Delta\varphi$ corresponding to the switch on-state to allow a fair comparison between the two device configurations. For MZIs, the phase shift can be straightforwardly defined by the condition:

$$\Delta\varphi = \frac{2\pi}{\lambda_0} \cdot \Delta n_{eff} L = \pi, \tag{X.2}$$

where $\Delta n_{eff}$ denotes the waveguide effective index change due to thermo-optic effect, $\lambda_0$ is the free space wavelength, and $L$ represents the physical length of the heated waveguide section. The waveguide effective index change is correlated to the temperature rise $\Delta T$ in the heating zone as:

$$\Delta n_{eff} = \Delta T \cdot \sum_j \left(\frac{dn}{dT}\right)_j \Gamma_j = \Delta T \cdot n_g \left(\frac{dn}{dT}\right)_{eff}. \tag{X.3}$$

In Eq. X.3, $dn/dT$ stands for the material thermo-optic coefficient and $\Gamma$ denotes the confinement factor following the definition by Robinson *et al.*[43]. The subscript $j$ labels different regions of the waveguide structure (e.g., core and cladding). It is worth noting that slow light effect, if any, is taken into account implicitly in the confinement factor (which can be much greater than unity when strong slow light effects present). $(dn/dT)_{eff}$ is the effective thermo-optic coefficient of the waveguide given by the weighted average of the thermo-optic coefficients of the waveguide's constituent materials:

$$\left(\frac{dn}{dT}\right)_{eff} = \sum_j \frac{\gamma_j}{n_j} \cdot \left(\frac{dn}{dT}\right)_j. \tag{X.4}$$

where $n$ is the material index and $\gamma$ denotes spatial confinement of the energy density following Ref. [43].

For resonant cavity switches, we similarly define the on-state as:

$$\Delta\varphi = \frac{2\pi}{\lambda_0} \cdot \Delta n_{eff} L_{eff} = \pi. \tag{X.5}$$

Here $L_{eff}$ is the effective propagation length in the cavity given by[44]:

$$L_{eff} = \frac{Q\lambda_0}{2\pi n_g}, \tag{X.6}$$

where $Q$ is the loaded cavity quality factor and $n_g$ denotes the modal group index. The thermo-optic resonance detuning corresponding to the phase shift $\Delta\varphi$ can be calculated by combining Eqs. X.5 and X.6:

$$\Delta\lambda = \frac{\Delta n_{eff}}{n_g} \cdot \lambda_0 = \frac{\pi\lambda_0}{Q}, \tag{X.7}$$

i.e. $\pi$ times FWHM of the resonant peak.

The temperature rise in the heating zone of an MZI switch in its on-state is then:

$$\Delta T = \frac{\lambda_0}{2n_g L} \Big/ \left(\frac{dn}{dT}\right)_{eff}, \tag{X.8}$$

and the corresponding power consumption is:

$$P_\varphi = \Delta T \cdot G = \frac{\lambda_0 G}{2n_g L} \Big/ \left(\frac{dn}{dT}\right)_{eff}. \tag{X.9}$$

In the case of resonant devices, we have:

$$P_\varphi = \frac{\pi G}{Q} \Big/ \left(\frac{dn}{dT}\right)_{eff}. \tag{X.10}$$

The figure of merit (FOM) for thermo-optic switches, often defined as inverse of the product of consumed power and 10%-to-90% rise time $t_r$ of the switch[45], becomes:

$$\frac{1}{P_\varphi \cdot t_r} = \frac{0.45}{P_\varphi \cdot \tau} = \frac{0.9 n_g L}{\lambda_0 C} \cdot \left(\frac{dn}{dT}\right)_{eff}, \tag{X.11}$$

in the case of MZI, and

$$\frac{1}{P_\varphi \cdot t_r} = \frac{0.45Q}{\pi C} \cdot \left(\frac{dn}{dT}\right)_{eff}. \tag{X.12}$$

for resonant switches. Here the 10%-to-90% rise time is trivially connected to the thermal time constant via:

$$t_r = 2.2\tau. \tag{X.13}$$

We note that some previous reports quoted the FOM of resonator-based thermo-optic switches with respect to the power consumption needed to shift the resonance by one free spectral range (FSR). This definition does not correctly represent the performance of switches, as the extinction ratio of a resonator switch is correlated to its resonance FWHM rather than FSR.

From Eqs. X.11 and X. 12 we can see that the FOM of MZIs scales with the group index whereas in resonant switches slow light effects have no impact on the FOM (note: there is a factor of $n_g$ in the expression of $Q$, although it is cancelled out as all linear losses are amplified by a factor of $n_g$ due to the slow light effect). Eq. X.12 suggests that the large FOM observed in our graphene thermo-optic switch benefits from both the strong thermal confinement (and hence a small heat capacity $C$) as well as the low parasitic optical absorption by graphene thanks to our sandwich waveguide design (which is conducive to a high $Q$).

Lastly, we derive the expression for the energy efficiency or heating efficiency $\eta$ (defined as the ratio of resonant wavelength shift over input power). The thermo-optic resonance shift $\Delta\lambda$ is[44]:

$$\Delta\lambda = \frac{\Delta n_{eff}}{n_g} \cdot \lambda_0 = \lambda_0 \Delta T \cdot \left( \frac{dn}{dT} \right)_{eff} , \tag{X.14}$$

which yields the energy efficiency as:

$$\eta = \frac{\Delta\lambda}{P_\varphi} = \frac{\lambda_0}{G} \cdot \left( \frac{dn}{dT} \right)_{eff} . \tag{X.15}$$

Therefore, the superior energy efficiency in our device is attributed to the low thermal conductance *G*, which again takes advantage of our unique device geometry where the graphene heater is embedded inside a glass waveguide with minimal contact area with the surrounding environment (and hence reduced thermal leakage)[46-48].

### Section X – Performance comparison of on-chip thermo-optic switches

Table S2 compares the performance of our photonic crystal thermo-optic switch with other on-chip thermo-optic switches reported in literature. Based on the derivation in Supplementary Section IX, the power $P_\varphi$ is taken as the power needed to induce a $\pi$ phase shift in an MZI arm or to detune a resonance by $\pi$ times FWHM of the resonant peak. For consistency, the table lists 10%-to-90% rise time $t_r$ of the switch devices (the rise time and fall time are identical in our device). We converted thermal time constant $\tau$ quoted in literature to $t_r$ using Eq. X.13 when applicable.

**Table S2. Performance summary of on-chip thermo-optic switches: cells with green shading correspond to results obtained with graphene heaters. "N/A" indicates that the result is not reported in literature and cannot be inferred from data presented.**

| Device | Energy efficiency (nm/mW) | Power $P_\varphi$ (mW) | Rise time $t_r$ (μs) | FOM = $\frac{1}{P_\varphi \cdot t_r}$ ($mW^{-1} \cdot \mu s^{-1}$) |
|---|---|---|---|---|
| Photonic crystal thermo-optic switch with embedded graphene heater (this report) | 10 | 0.11 | 14 | 0.65 |
| Graphene heater on slow light photonic crystal MZI[49] | 1.07 | 2.0 | 0.75 | 0.67 |
| Graphene heater on micro-disk resonator[41] | 1.7 | 14 | 12.8 | 0.0056 |
| Graphene heater on micro-ring resonator[50] | 0.33 | 1.9 | 3 | 0.18 |
| Graphene heater on micro-ring resonator[40] | 0.10 | 120 | 0.75 | 0.011 |
| Doped silicon in slow light photonic crystal MZI[51] | 0.7 | 2 | 0.19 | 2.6 |
| Doped silicon heater embedded in micro-ring resonator[52] | 0.12 | 3.7 | 0.84 | 0.32 |
| MZI with metal heaters and a graphene heat conductor[53] | 0.06 | 70 | 20 | 0.0007 |
| NiSi heaters integrated with Si waveguides[54] | N/A | 20 | 6.2 | 0.008 |
| Doped silicon heater embedded in MZI[45] | N/A | 12.7 | 5.3 | 0.015 |
| Doped silicon heater embedded in MZI[55] | 0.25 | 24.8 | 2.7 | 0.015 |
| Metal heater on suspended MZI[56] | N/A | 0.54 | 141 | 0.013 |
| Metal heater on suspended Michelson interferometer[57] | N/A | 0.05 | 780 | 0.026 |
| Metal heater on Si waveguide Fabry-Perot cavity[58] | 0.045 | 24.4 | 1.9 | 0.022 |
| Metal heater on MMI[59] | N/A | 24.9 | 2.6 | 0.015 |

We see from Table S2 that our device claims the highest energy efficiency so far among all on-chip thermo-optic switches. Our device also achieves a low switching power without resorting to suspended structures, and a high FOM on par with the best values from prior reports. In most cases, our device's FOM is orders of magnitude better than those measured in thermo-optic switches without slow light enhancement.

**Section XI – Modeling of wavelength-dependent absorption in graphene detector**

Wavelength dependent absorption of graphene in the mid-IR photodetector was modeled using the software MODE Solutions (Lumerical Inc.). In the simulation, graphene was treated as a layer with zero thickness based on the surface conductive model[60], and its Fermi level is set to be 0.34 eV below the Dirac point according to our Hall measurement. Figure S19a depicts the waveguide TE mode profile overlaid on an SEM image showing the fabricated ChG-on-graphene waveguide mid-IR detector. The waveguide dimensions are as follows: ridge width 1.3 μm, slab thickness 0.25 μm and ridge height 0.5 μm. The calculated wavelength-dependent graphene absorption is plotted in Fig. S19b. The observed trend manifests the combination of two effects: reduced graphene absorption at longer wavelength due to the onset of Pauli blocking, and increased modal field overlap with the graphene layer at longer wavelength. The former effect dominates the wavelength dependence shown in Fig. S19b.

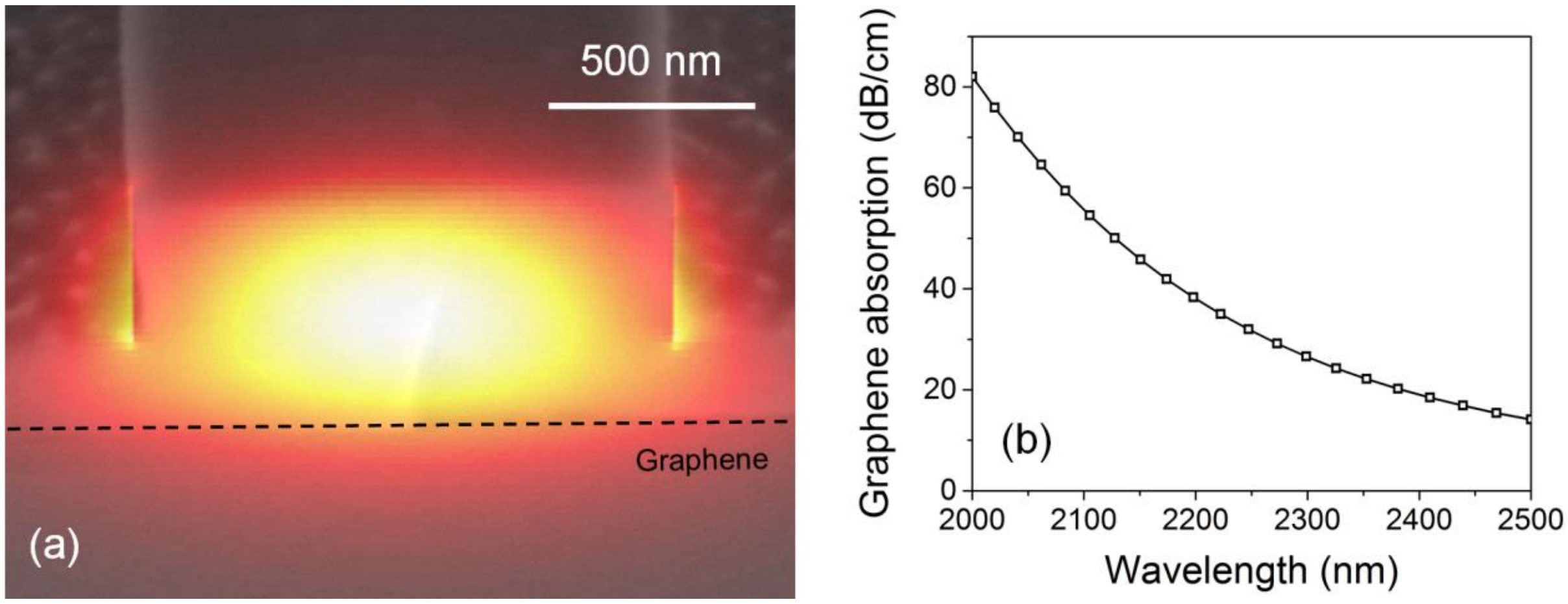

**Fig. S19. (a) TE mode profile of the mid-IR waveguide-integrated detector at 2 μm wavelength overlaid with an SEM image of the fabricated device, where the dotted line marks the location of the graphene layer; (b) simulated TE mode optical absorption in the graphene layer as a function of wavelength.**

**Section XII – Fabrication and characterization of flexible waveguide-integrated graphene detectors**

Using our integration scheme, we demonstrated, for the first time to the best of our knowledge, waveguide-integrated graphene photodetectors on a flexible polymer substrate. Fabrication protocols of the detector are similar to that used for mid-IR detector processing described in Methods, except that we used a thermally oxidized silicon wafer coated with an SU-8 epoxy layer as the handler substrate onto which graphene is transferred and the detectors were fabricated. The SU-8 membrane can be delaminated from the handler substrate after detector fabrication to form free-standing flexible devices. This process, which has previously been adopted for flexible ChG photonic device fabrication by the authors[61-65], capitalizes on the low deposition and processing temperatures of ChG's to facilitate direct integration on polymers which typically cannot withstand temperatures above 250 °C. The $Ge_{23}Sb_7S_{70}$ glass waveguide dimensions are as follows: ridge width 0.8 μm, slab thickness 0.13 μm and ridge height 0.3 μm.

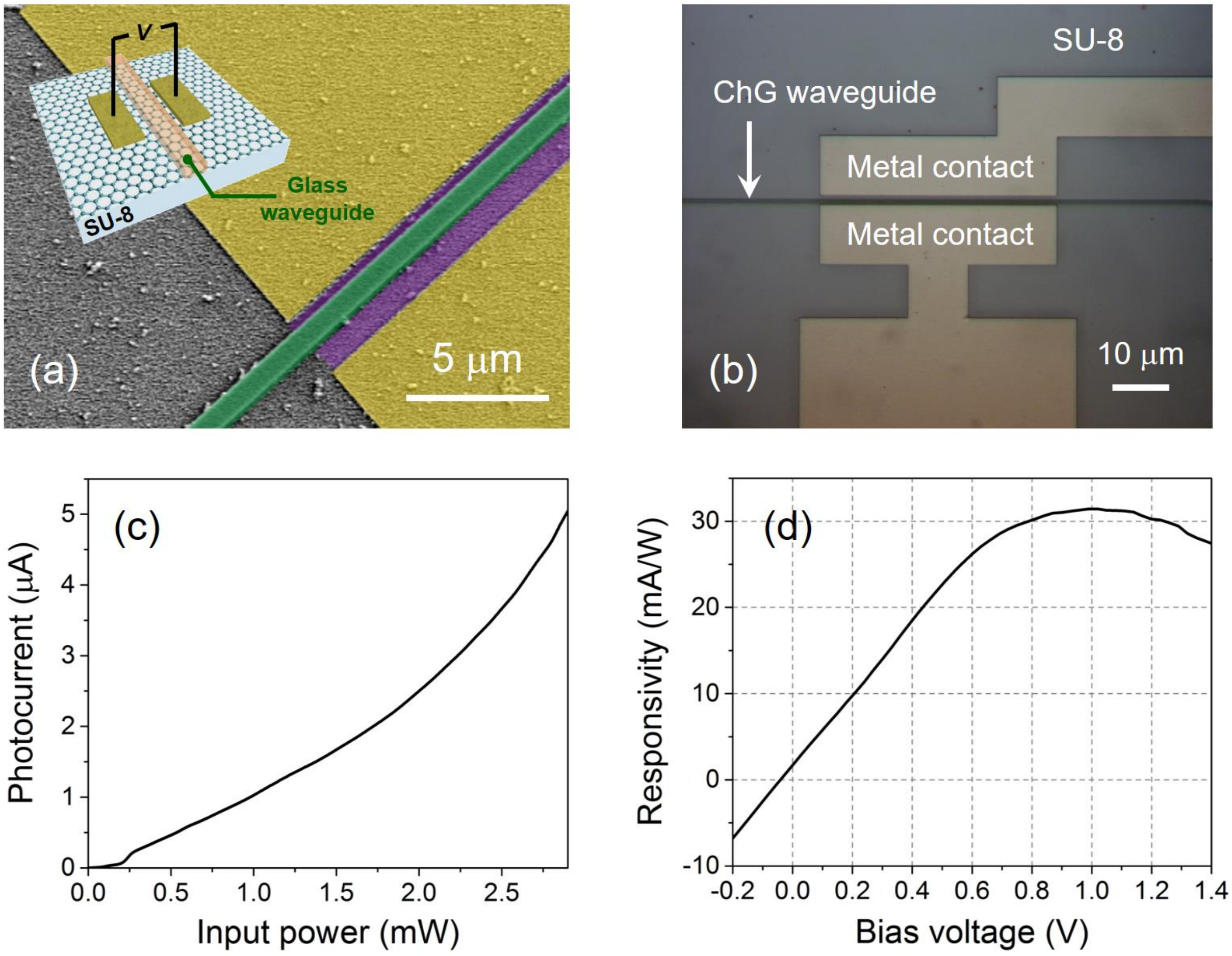

**Fig. S20. (a) False color SEM tilted view of the flexible waveguide-integrated photodetector; inset schematically illustrates the detector structure; (b) top-view optical micrograph of the detector; (c) photocurrent measured at zero bias as a function of input optical power at 1550 nm wavelength from the waveguide; (d) bias-voltage dependent responsivity of the detector at 1550 nm.**

Figures S20a and S20b display images of the fabricated detector device. It is apparent from Fig. S20a that the waveguide is situated closer to one metal electrode than to the other to trigger non-vanishing photothermoelectric (PTE) response. We have also fabricated and tested devices where the waveguide is positioned at equal distance from the two metal electrodes and assumes an otherwise identical configuration. As expected, we measured no photoresponse in such devices, which corroborates the PTE device operation mechanism.

The detector was measured at a flat (i.e. without deformation) state following protocols identical to mid-IR detector measurement albeit at 1550 nm wavelength. Figure S20c plots the magnitude of photocurrent measured at zero bias as a function of the guided optical power in the feeding waveguide. Figure S20d shows the detector's responsivity dependence on bias voltage: its responsivity reaches a maximum value of 32 mA/W at 1 V bias and decreases at higher bias voltage. Such a response is characteristic of graphene photodetectors operating in the PTE mode[66]. We attribute the reduced responsivity (compared to the peak responsivity of 250 mA/W measured in the mid-IR detector) to increased series resistance. The total electrical resistance of the flexible detector is 1650 ohm, much higher than that measured in the mid-IR detector device (260 ohm). Both detectors have the same length (80 μm) and the spacing between the metal contacts is 1.8 μm for the flexible detector and 2.5 μm for the mid-IR detector. Since the sheet resistance of our CVD graphene is approximately 400 ohm/sq., the large difference in resistance cannot be accounted for by the different device geometries and is likely a consequence of increased contact resistance resulting from unoptimized device processing. Further streamlined device fabrication process is anticipated to significantly boost the flexible detector performance.

## Section XIII – Graphene modulator modeling

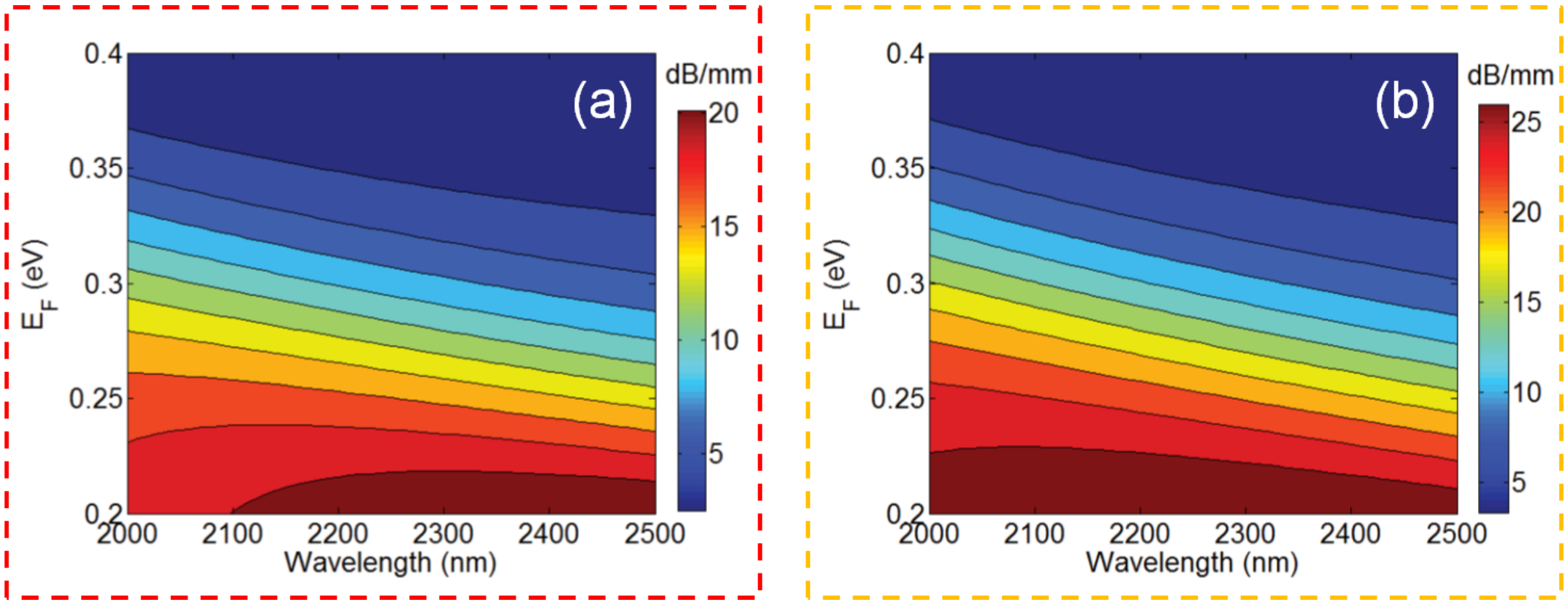

**Fig. S21. Calculated modulator waveguide loss contribution due to optical absorption in the (a) bottom; and (b) top layer graphene as functions of graphene Fermi level. The total optical attenuation in the modulator device is the sum of the two absorption loss terms.**

The graphene mid-IR waveguide modulator was modeled by quantifying the voltage-dependent optical absorption in the two graphene layers. Using the surface conductive model[60] in conjunction with a waveguide mode solver (MODE Solutions, Lumerical Inc.), we first calculated the modulator waveguide loss incurred by optical absorption of the two graphene layers embedded in the device, and the results are graphically represented in Figs. S21a and S21b. Once Fermi levels of the two graphene layers are known, the total optical attenuation through the modulator device can be calculated by summing up the loss contributions from the two graphene layers. The next step is to solve the voltage-dependent Fermi levels in graphene. We start with the expression of density of states $g$ in monolayer graphene[67]:

$$g(E) = \frac{2}{\pi(\hbar V_F)^2} \cdot |E|, \qquad \text{(XIII.1)}$$

where $E$ represents energy relative to the Dirac point, $\hbar$ is the reduced Planck constant, and $V_F \sim 10^6$ m/s gives the Fermi velocity of carrier in graphene. Integrating Eq. XIII.1 with respect to energy, we obtain the carrier density $n_c$ in graphene as a function of Fermi level $E_F$:

$$n_c = \frac{1}{\pi(\hbar V_F)^2} \cdot E_F{}^2, \qquad \text{(XIII.2)}$$

In the absence of applied bias, the Fermi levels in the two graphene layers in the modulator are 0.355 eV (bottom layer) and 0.325 eV (top layer) below the Dirac point, which define the initial carrier concentrations in the two graphene sheets according to Eq. XIII.2. When a bias voltage is applied across the two graphene layers, charge is transferred from one layer to the other to build up the electric potential between them. Modeling the two-layer system as a simple parallel plate capacitor, the number of charge carriers transferred at a bias voltage $V$ is:

$$\Delta n_c = \frac{\varepsilon \varepsilon_0 V}{d}, \qquad \text{(XIII.3)}$$

where $\varepsilon$ denotes relative permittivity (dielectric constant) of the ChG gate dielectric, $\varepsilon_0$ symbolizes the permittivity of vacuum, and $d$ is the spacing between the two graphene layers (i.e. gate thickness). The carrier densities in the two graphene layers under the bias $V$ are given by $n_c + \Delta n_c$ and $n_c - \Delta n_c$, respectively. The Fermi levels in the two graphene layers are then solved from Eq. XIII.2 and used as input parameters to calculate the waveguide loss induced by graphene absorption using the data shown in Fig. S21.

The above outlines the procedures to quantify modulator insertion loss at a given bias voltage. In the calculation, the gate thickness $d$ in our device is measured from the SEM image shown in Fig. 5b to be 50 nm. The dielectric constant $\varepsilon$ of $Ge_{23}Sb_7S_{70}$ glass is left as a fitting parameter. Based on the measurement results in Fig. 5c, we find that $\varepsilon = 3.0$ provides the best fit and the corresponding modeling results are plotted in Fig. 5d.

**Section XIV – Bandwidth of graphene modulators: analysis and performance projection**

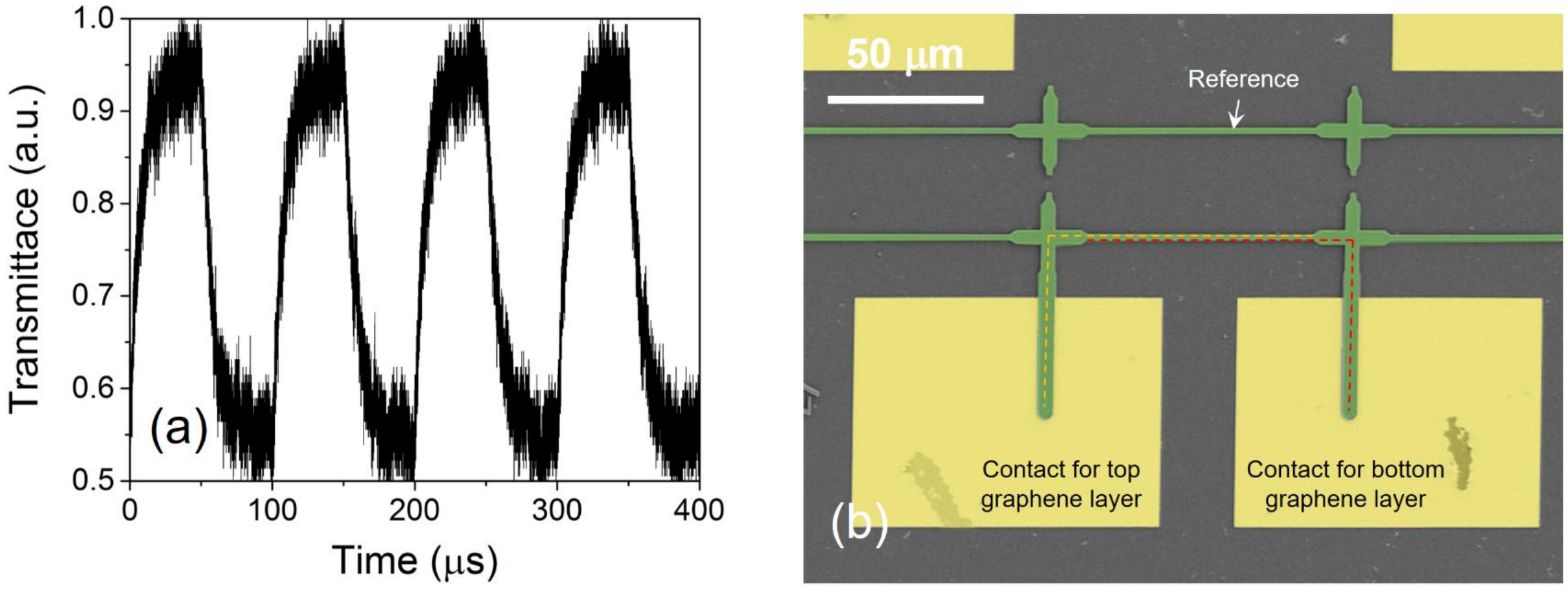

**Fig. S22. (a) Measured time-domain response of the mid-IR graphene modulator to a square wave driving voltage input modulated at 10 kHz; (b) false color top-view SEM micrograph showing a mid-IR graphene modulator device. The scratches on the metal electrodes were inflicted by electrical contact probes during measurement. The dotted lines represent the AC current paths from the contacts to the two graphene layers. The image also shows a reference device on top which is not electrically connected and was only used for optical testing.**

Bandwidth of the mid-IR graphene modulator device was quantified by monitoring its time-domain response to a square wave driving voltage input at 10 kHz (Fig. S22a). From the figure, we fit the time constant of the modulator to be $\tau$ = 7 μs. The 3-dB bandwidth of the device is inferred via:

$$f_{3dB} = \frac{1}{2\pi\tau} \tag{XIV.1}$$

to be 23 kHz.

The relatively slow response of the modulator is attributed to the large capacitance of electrical probes used in our test and series resistance from the graphene layers. Figure S22b shows an SEM top-view image of a 100-μm-long modulator device showing both contacting electrodes for the two graphene layers. Similar to the thermo-optic switch, the embedded graphene layers are electrically connected to the metal electrodes via a waveguide crossing structure. Unlike the thermo-optic switch where the two electrodes are electrically linked through the graphene layer, each of the two graphene layers is individually connected to an electrode to form a parallel plate capacitor between the graphene layers. The 700-μm-long device on which we performed the modulation and bandwidth measurements has an identical configuration except that the spacing between the two waveguide crossings are larger. This configuration, however, leads to large series resistance due to the long current path in the graphene strips embedded in the waveguide. Based on the measured sheet resistance value of 400 ohm/sq. in our CVD graphene and the device geometry, the calculated series resistance due to graphene is $4.3 \times 10^5$ Ω. The modulator capacitance is estimated using a simple parallel plate capacitor model to be about 0.5 pF for a 700-μm-long device. In addition, the electrical contact probes we used in the bandwidth measurement were not designed for high-speed tests and electrical capacitance associated with the probes were assessed using an RLC meter (Agilent 4284A Precision LCR Meter) to be ~ 20 pF. The RC time constant defined by the parasitic probe capacitance and the graphene series resistance is ~ 8.6 μs, which agrees with our measured result of 7 μs.

The RC-limited bandwidth of the two-layer graphene modulator can be significantly boosted if a side-contact scheme is implemented, where each graphene layer extends outside one side of the waveguide and is electrically contacted[68,69]. Using the same graphene sheet resistance value of 400 ohm/sq., the series resistance of the graphene sheet can be readily reduced to less than 10 Ω. State-of-the-art graphene optoelectronic devices also claim contact resistance between metal electrodes and graphene sheets less than 100 Ω·μm[70-72], corresponding to less than a few Ohms of resistance in the case of the side-contact geometry. Therefore, assuming a 50 Ω load and the same device capacitance of 0.5 pF as that of our modulator, an RC-limited 3-dB bandwidth of 6.4 GHz is projected. Indeed, high-speed modulation at 1 GHz has been experimentally demonstrated in a side-contacted graphene modulator operating near the 1550 nm telecommunication wavelength[68].

## Section XV – Mid-infrared measurement system

The measurement system used to test and characterize the graphene mid-IR modulator and detector devices uses a 2.0 – 2.55 μm wavelength tunable CW laser (IPG Photonics Co.). The Cr:ZnSe/S gain medium is pumped by a high-power near-infrared fiber laser and emits up to 1.5 W of power in the mid-infrared. Wavelength accuracy was ensured by calibrating the laser controller with a mid-infrared optical spectrum analyzer (Thorlabs OSA207). For measurement of both devices, the output light was chopped at approximately 1.5 kHz, and the output was detected via a lock-in amplifier. A $CaF_2$ wedge re-directed about 3-4% of the light to an InGaAs photodetector (Thorlabs DET10D) so that the output can be appropriately normalized by the input power. After this, the light passes through a chalcogenide molded aspheric lens and is coupled into the single-mode on-chip waveguides. To approximate the coupling losses from converting the free-space optical mode to the waveguide mode, we aligned two identical chalcogenide aspheric lenses to the input and output of a sample with only a single-mode waveguide. We measured the optical power before the input asphere (labelled as P1 in the figures) and after the output asphere (not shown) using a thermal power meter (Thorlabs S302C), then divided the loss in dB by two to obtain the coupling loss per device facet.

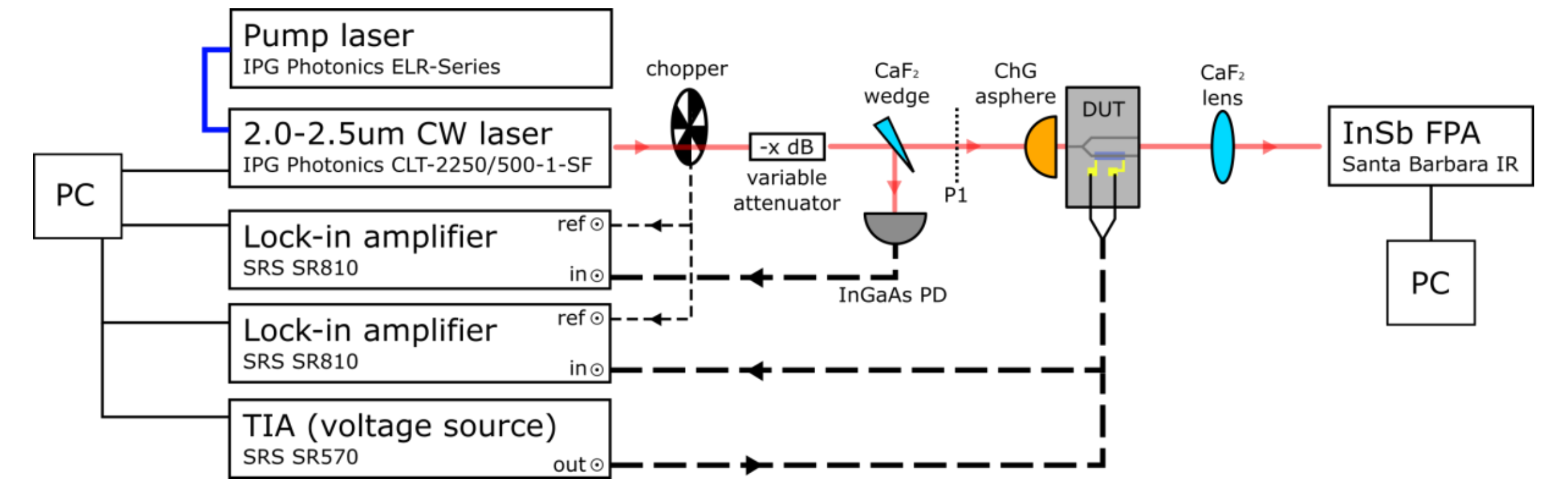

**Fig. S23. Block diagram illustrating the measurement system configured to characterize the mid-infrared graphene photodetector**

To measure the waveguide integrated graphene photodetector, the device was first optically aligned with a microscope objective above the sample and an InSb focal plane array (Santa Barbara Infrared, Inc.) at the output, as shown in Fig. S23. Once optically aligned, the absolute value of the photocurrent was measured without chopping. Next, the light was chopped and the signal was recorded via a lock-in amplifier. This procedure allowed us to map the voltage recorded at the lock-in amplifier to the device's corresponding DC photocurrent.

Linearity of the photodetector was characterized by adjusting the optical power with a variable attenuator, measuring the optical power at P1, and measuring the device photocurrent. The responsivity of the photodetector was determined as a function of both wavelength and applied detector bias at fixed input power. The current preamplifier's output bias was first swept from -1.5 V to +1.5 V bias at a fixed wavelength and power. Finally, the laser wavelength was swept between 2.0 and 2.5 μm wavelength at zero bias and fixed power. In all measurements, the signal was appropriately normalized by the corresponding input power, as determined by the InGaAs reference detector.

The graphene electroabsorption modulator was characterized in the mid-IR by a similar setup, although modified since the output signal of interest is the intensity of light after the modulator. This light is out-coupled from the device and refocused onto a second InGaAs photodetector (Thorlabs DET10D) with an iris to subtract any light not originating from the waveguide mode, as shown in Fig. S24. The modulator was biased using a DAQ module (NI USB-6212) and the output optical signal measured by a lock-in amplifier (SRS SR810). In a similar manner, the modulator was characterized as a function of wavelength and voltage bias. Finally, the dynamic response of the electroabsorption modulator was tested by modulating the bias rather than the input optical signal and measuring the temporal response with an oscilloscope (Tektronix DPO 2014), as illustrated in Fig. S25.

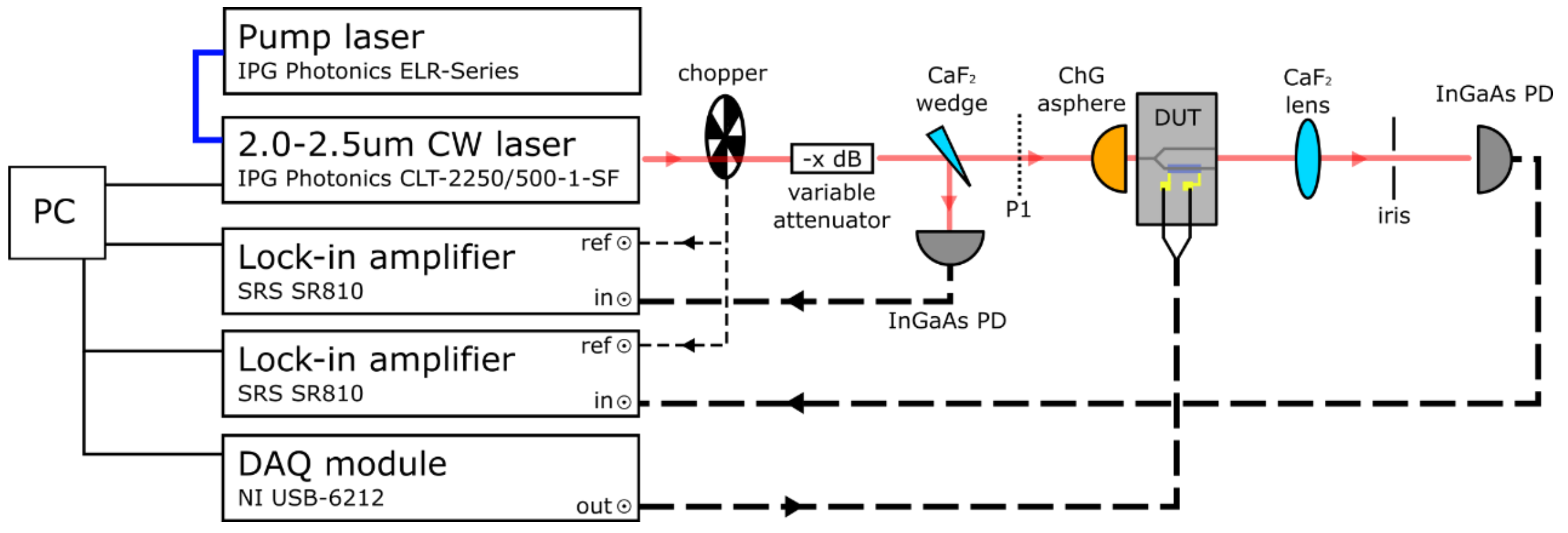

**Fig. S24. Block diagram illustrating the measurement system configured to characterize the mid-infrared graphene electroabsorption modulator**

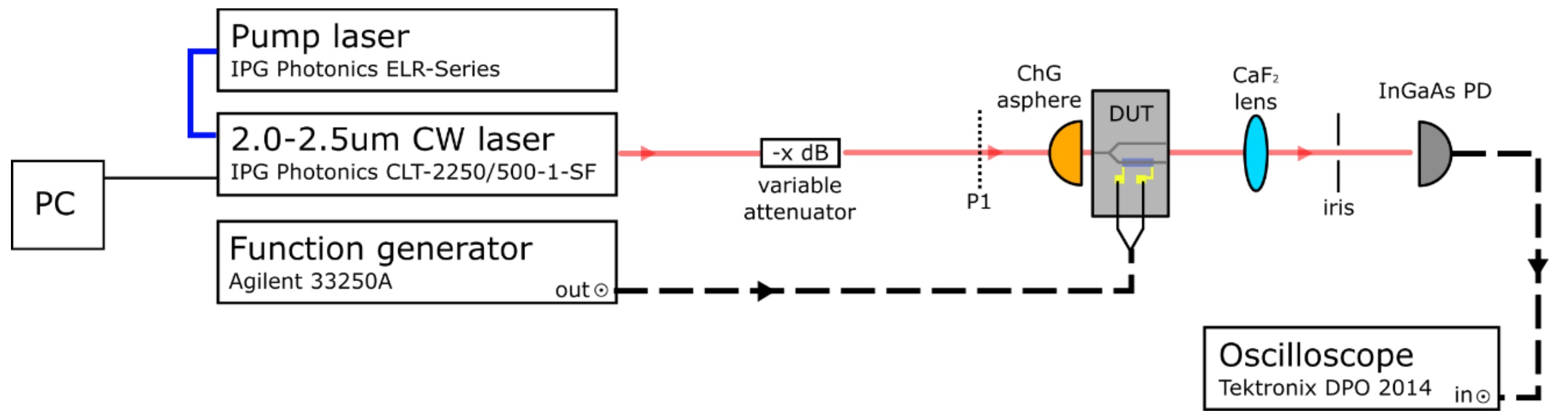

**Fig. S25. Block diagram illustrating the measurement system configured for speed testing of the mid-infrared graphene electroabsorption modulator**